\documentclass[journal]{IEEEtran}
\usepackage{amsmath,amsfonts}
\usepackage{algorithmic}
\usepackage{algorithm}
\usepackage{array}
\usepackage[caption=false,font=normalsize,labelfont=sf,textfont=sf]{subfig}
\usepackage{textcomp}
\usepackage{stfloats}
\usepackage{url}
\usepackage{verbatim}
\usepackage{graphicx}
\usepackage{cite}
\usepackage{booktabs}
\usepackage{tabularx}
\usepackage{siunitx}
\usepackage{multirow}
\usepackage{subcaption}
\usepackage{makebox}
\usepackage{supertabular}
\usepackage{lipsum}
\usepackage{adjustbox}
\usepackage{dcolumn}
\usepackage{amssymb}
\usepackage{pdfpages}
\usepackage{array}
\usepackage{siunitx}
\usepackage{makecell}

\newcommand{\mytoprule}{
	\specialrule{.1em}{0pt}{0pt}
	\rule{0pt}{1em}
}

\newcommand{\mymidrule}{
	\hline
	\hline
}

\newcommand{\mybottomrule}{
	\specialrule{.1em}{0pt}{0pt}
}

\newcommand{\mylineheight}{
	\renewcommand\arraystretch{1.1	}
}

\begin{document}

\title{Measuring the Sense of Presence and Learning Efficacy in Immersive Virtual Assembly Training}

\author{Weichao Lin, Liang Chen, Wei Xiong, Kang Ran, Anlan Fan
  \thanks{Manuscript received MM DD, 2023; revised MM DD, 2023.}
  }

\markboth{Journal of \LaTeX\ Class Files,~Vol.~14, No.~8, August~2021}%
{Shell \MakeLowercase{\textit{et al.}}: A Sample Article Using IEEEtran.cls for IEEE Journals}

\IEEEpubid{0000--0000/00\$00.00~
	}

\maketitle

\begin{abstract}
With the rapid progress in virtual reality (VR) technology, the scope of VR applications has greatly expanded across various domains. However, the superiority of VR training over traditional methods and its impact on learning efficacy are still uncertain. To investigate whether VR training is more effective than traditional methods, we designed virtual training systems for mechanical assembly on both VR and desktop platforms, subsequently conducting pre-test and post-test experiments. A cohort of 53 students, all enrolled in engineering drawing course without prior knowledge distinctions, was randomly divided into three groups: physical training, desktop virtual training, and immersive VR training. Our investigation utilized analysis of covariance (ANCOVA) to examine the differences in post-test scores among the three groups while controlling for pre-test scores. The group that received VR training showed the highest scores on the post-test. Another facet of our study delved into the presence of the virtual system. We developed a specialized scale to assess this aspect for our research objectives. Our findings indicate that VR training can enhance the sense of presence, particularly in terms of sensory factors and realism factors. Moreover, correlation analysis uncovers connections between the various dimensions of presence. This study confirms that using VR training can improve learning efficacy and the presence in the context of mechanical assembly, surpassing traditional training methods. Furthermore, it provides empirical evidence supporting the integration of VR technology in higher education and engineering training. This serves as a reference for the practical application of VR technology in different fields.
\end{abstract}


\section{Introduction}

\IEEEPARstart{W}{ith}  the advent of the metaverse concept and the continuous iteration of interactive hardware in recent years, virtual reality (VR) and augmented reality (AR) are both evolving at a pace never before seen in the past and are entering the minds of consumers, not just in the lab. VR technology has been successfully integrated into various fields, including education, training, entertainment, and tourism.

The application of virtual reality in various industry training scenarios is gaining popularity, such as in mechanical\cite{huoDesignSimulationVehicle2022}, architectural\cite{liSynthesizingPersonalizedConstruction2022}, banking\cite{kickmeier-rustVirtualRealityProfessional2020}, firefighting\cite{jeonMoreBetterImproving2021}, medical\cite{kennedyImprovingSafetyOutcomes2023,luInnovativeVirtualReality2023,pedramValidationVRHMDsMedical2023,serranovergelComparativeEvaluationVirtual2020}, military\cite{kullmanVRMRSupporting2019}, and more. Immerse's report suggests that VR training programs can positively influence learners' attitudes, increasing their willingness to learn and experiment\cite{ImmerseSaysVR2021}. While most studies report higher user satisfaction with VR compared to traditional teaching methods, there is limited research addressing factors like presence and efficacy. To ensure successful VR experiences, key factors directly related to user experience must be taken into account, and their correlation with learning efficacy measured\cite{checaImmersiveVirtualrealityComputerassembly2021}.Understanding these elements' impact on learning efficacy remains a research gap.


\IEEEpubidadjcol

Engineering Drawing, a fundamental technical course, aims to cultivate students' abilities in reading and creating engineering drawings, with a particular focus on the training of spatial cognition in mechanical structures, which constitutes a central and challenging aspect of the curriculum. However, in mainland China, traditional teaching methods, relying on textbooks and presentations, fail to effectively convey structural aspects and engage learners, leading to reduced interest and learning efficacy. While physical training remains effective, it is constrained by the limited availability of school model resources and the associated maintenance costs, making its widespread promotion challenging.


In this study, for mechanical structure assembly training, three distinct methods were compared: physical model training, desktop program simulation training, and immersive VR simulation training.  The comparison is based on a classic training task in engineering drawing courses, which involves assembling a bench vise.  The desktop simulation training programs and VR simulation training programs were developed specifically for this study and have been specially customized to suit the experimental equipment used.

The study aims to assess VR training's effectiveness by comparing it with two other methods. Conducting experiments on a wide range of undergraduate students to compare three different learning methods is a prerequisite for ensuring the validity of the conclusions. The sense of presence in VR depends on users shifting their focus from the physical environment to the virtual one. The degree to which users focus on the virtual environment to some extent determines their level of engagement in that environment. However, the relationship between this sense of presence and training efficacy remains unclear. This study investigates whether there are differences in presence among different systems, the correlation of various factors, and their potential impact on learning efficacy, using Witmer's five factors for measuring presence in virtual environments\cite{witmerMeasuringPresenceVirtual1998}.

\IEEEpubidadjcol



This study represents a pioneering exploration of novel pedagogical methods in higher education, and the results of our analysis will assist educational teams in higher institutions in the development of advanced immersive simulation-based assembly training programs. This, in turn, will provide critical guidance for enhancing the quality of undergraduate education. Furthermore, these research outcomes will also offer valuable support for training initiatives within the manufacturing industry, contributing to the enhancement of assembly skills and efficiency among employees, ultimately bolstering the competitiveness of enterprises. The following sections are structured as follows: Section 2 reviews relevant literature on VR training. Section 3 outlines the development of immersive VR simulation training for engineering drawing teaching. Section 4 describes the experimental process. Section 5 presents experimental results. Section 6 discusses the evaluation and Section 7 concludes the study and outlines future research directions.


\section{Related work}

\subsection{Applications of VR in Medicine, Architecture, Civil Engineering, and Mechanical Engineering}

Virtual reality technology has been widely employed across various domains for educational and training purposes, including medical surgical training, architectural and civil engineering practice training, and mechanical assembly instruction. Immersive technologies present themselves as conduits for rich, profoundly interactive, captivating, and secure learning experiences within the realm of medical practice and education, emphasizing the seamless transfer of skills to real clinical environments\cite{buttUsingGameBasedVirtual2018}. Research has particularly focused on simulation-based training, with studies showcasing the capacity of immersive technology to augment students' acquisition of knowledge\cite{sultanExperimentalStudyUsefulness2019,zhaoEffectivenessVirtualRealitybased2020}. In the context of minimally invasive surgery, the deployment of VR simulators has demonstrated marked enhancements in the performance scores of participants\cite{guedesVirtualRealitySimulator2019}.

Within the field of architecture, students necessitate a heightened spatial imagination. However, the conventional presentation of architectural drawings in two dimensions imposes limitations on students' comprehension of application concepts\cite{kandiApplicationVirtualReality2020}. To overcome this challenge, scholars have proposed immersive VR systems that guide students in comprehending the sequential construction processes of bridges\cite{sampaioApplicationVirtualReality2014}. Such systems facilitate the visualization of structures and foster interactive information exchange pertaining to their physical behavior. Additionally, research has delved into the application of VR systems to facilitate experimental practices in architecture, such as employing VR simulations to guide students through concrete compression experiments\cite{vergaraNewApproachTeaching2017}. These virtual reality models furnish fresh perspectives in pedagogy, fostering interactive engagement between students and teachers with virtual models, thereby contributing to discussions on novel subjects and complex construction sequences.

In the realm of civil engineering, structural analysis serves as a cornerstone of knowledge. However, due to its abstract concepts and intricate algorithms, teaching structural analysis often poses challenges\cite{chouConstructionVirtualReality1997}. A study named SISMILAB explores the merits of virtual experiments in comprehending fundamental concepts related to seismic engineering\cite{gomezDevelopmentVirtualEarthquake2018}. In parallel, to enhance the efficacy of practical training, researchers have proposed virtual training systems aimed at improving operators' control skills in the operation of construction excavators\cite{mastliInteractiveHighwayConstruction2017}.

In the field of mechanical engineering, assembly systems represent dynamic processes wherein components are systematically interconnected to realize the desired final product. Since the assembly of any product consumes additional time and incurs costs throughout the manufacturing process, assembly workers should possess a comprehensive understanding of the assembly sequence and the requisite tools\cite{eswaranChallengesOpportunitiesAR2022}. Adas et al. have presented a learning environment that leverages the potential of virtual reality and augmented reality (VAR) to immerse students in authentic assembly design challenges, allowing for interactive engagement with both virtual and tangible mechanical components in a guided and interactive manner\cite{adasVirtualAugmentedReality2013}. Winkes et al. have conducted a systematic evaluation of assembly decisions and training for assembly operations using a virtual reality workshop equipped with comprehensive virtual reality technology. The functionalities of the VR workshop encompass virtual representation, evaluation, and the formulation of improvement plans for assembly objects\cite{winkesMethodEnhancedAssembly2015}. Ajay Karthic et al. have conducted a comparative analysis of VR devices against desktop solutions to ascertain their efficacy in facilitating functional analysis of products, encompassing the assembly of coffee machine components\cite{bharathiInvestigatingImpactInteractive2016}. Moreover, Al-Ahmari et al. have developed a fully functional virtual manufacturing assembly simulation system, offering an interactive workstation for evaluating assembly decisions and providing training for assembly operations. This system is capable of delivering multi-channel interactive feedback, incorporating visual, auditory, and tactile sensations. The system boasts commendable completion and compatibility rates, rendering invaluable support for large-scale assemblies\cite{al-ahmariDevelopmentVirtualManufacturing2016}.

\subsection{Advantages of VR Training}

Virtual reality presents numerous advantages in the field of education when compared to traditional teaching methods\cite{zhaoEffectivenessVirtualRealitybased2020}. Firstly, VR has the capability to simulate rigorous and realistic environments or scenarios that may be challenging to replicate with current resources\cite{bennettMixedRealityPedagogicalInnovation2021,vergaraNewApproachTeaching2017,abidiAssessmentVirtualRealitybased2019}, offering cost-effective solutions \cite{hafsiaVirtualRealitySimulator2018,cassolaImmersiveAuthoringVirtual2021}. Secondly, in comparison to conventional training schemes or devices, such as desktop-based applications, VR demonstrates significant enhancements in information delivery and visualization \cite{gomezDevelopmentVirtualEarthquake2018}. Traditional experimental equipment exhibits greater complexity in terms of feedback and configuration \cite{tergasPilotStudySurgical2013}. Additionally, in contrast to mobile devices like smartphones and tablets, virtual reality mitigates the issue of screen glare under intense lighting conditions and eliminates the need to occupy one hand for device handling, while significantly expanding the visual field \cite{setarehApplicationVirtualEnvironment2005}.

Virtual reality technology provides students with highly immersive learning experiences, fostering their motivation and interest in learning \cite{buttUsingGameBasedVirtual2018}. Researchers have found that immersive technology training creates richly interactive and captivating educational environments, effectively enhancing student interest and intrinsic motivation for learning \cite{tergasPilotStudySurgical2013,wangTaskComplexityLearning2020}. Furthermore, virtual reality technology can offer higher fidelity and improve training outcomes \cite{becerik-gerberBIMEnabledVirtualCollaborative2012,mcmahanEvaluatingDisplayFidelity2012}. Immersive hardware, such as headsets and specialized input devices, is believed to yield superior training effects by enhancing the sense of physical and psychological realism \cite{duboviNowKnowHow2017,mogliaSystematicReviewVirtual2016,vaughanReviewVirtualReality2016}. Positive results have also emerged concerning the enhancement of spatial thinking and spatial imagination abilities through virtual reality training \cite{tuzunEffects3DMultiuser2016}.

The sensation experienced within virtual reality environments can be described as \textit{presence}. Research has indicated a close association between presence and attention, motivation, learning, and subsequent behavioral changes. For instance, when employing interactive digital environments with virtual orientation, learners may feel profoundly immersed within the digital realm, effectively mapping their experiences within the digital environment to the real world, thereby reducing the need for cumbersome "translation" between training and transfer environments \cite{tuzunEffects3DMultiuser2016}. Even in non-immersive environments, such as desktop-based systems, potent sensations of presence can be engendered in certain aspects \cite{duboviNowKnowHow2017}.Currently, there have been studies that utilize spatial presence to guide the design of virtual learning environments\cite{riegerHowMaximiseSpatial2023b}.

\subsection{Evaluating the Learning Efficacy}

Assessment of educational virtual reality (VR) training typically utilizes quantitative methods, such as pre- and post-test designs, in conjunction with questionnaires to measure learning outcomes\cite{xieReviewVirtualReality2021}. However, these approaches have limitations, especially when dealing with tasks involving intricate complex motion processes, as they may struggle to replicate real-world scenarios. Therefore, the evaluation of immersive VR assembly training must consider additional factors.


Koumaditis et al. conducted a comparative study on the efficacy of virtual training and physical training for teaching manual assembly tasks, introducing task complexity (TCXB) as an indicator of assembly errors in the final assembly process\cite{koumaditisEffectivenessVirtualPhysical2020}.

Checa et al. developed a VR serious game for training computer host assembly and compared it with other online learning methods, including online traditional lectures and playing the same serious game on a desktop computer. The results revealed that students exhibited higher satisfaction and engagement when utilizing the VR serious game. Covariance analysis was employed to examine the learning effects in this study\cite{checaImmersiveVirtualrealityComputerassembly2021}.

Winther et al. conducted a case study at Grundfos, a Danish pump manufacturer, comparing three training methods for teaching machine operation to new employees: pair-wise training, video training and VR training. Pair-wise training involved coaching sessions, while VR training took place in a dedicated room where trainees independently utilized VR devices for training. The study aimed to evaluate the efficacy of VR training for metric pump maintenance tasks compared to traditional training methods (video training and pair-wise training). The findings indicated that VR training was effective for instructional tasks, although both pair-wise training and video training outperformed VR training significantly\cite{wintherDesignEvaluationVR2020}.



While past research has implemented pre-test and post-test experiments, it is noteworthy that many studies have predominantly assessed system usability using subjective metrics such as user acceptance and satisfaction. Quantitative assessments of learning outcomes remain relatively scarce. In light of this, the present study further incorporates Witmer's five-factor scale for measuring presence in a virtual environment as an adjunct means of evaluation\cite{witmerMeasuringPresenceVirtual1998}.


\section{Creating Virtual Experiences for Assembly Training}

\subsection{Use Case Description}
Structural cognition training is challenging in engineering drawing education. Traditional offline classroom methods using textbooks, exercise books, and presentations lack intuitive portrayal of structures and fail to engage learners effectively, affecting learning efficacy. Physical training yields acceptable results but suffers from limited model availability and high maintenance costs, limiting widespread adoption. Leveraging virtual reality (VR) technology for structural cognition training can provide immersive and interactive learning environments, fostering practical skills and addressing traditional learning shortcomings\cite{chenApplicationAugmentedReality2011}.

This study targets the needs of learners and educators, drawing inspiration from constructivist learning theory. It focuses on the "Engineering Drawing" course for engineering students in higher education, using a bench vise for practice. The objective is to enhance learners' structural cognition, deepen their understanding of principles, and improve assembly-disassembly skills for mechanical products. The study designs training systems for two platforms: VR and desktop. A comparative analysis evaluates the learning outcomes of VR training, desktop training, and traditional physical training. The findings offer insights for the application and advancement of virtual reality technology in education.

\subsection{Theory and Design}

Within the realm of education, the incorporation of theoretical foundations is paramount for effective instructional design and implementation. To delve into the design and application of a virtual reality-based structural cognition training system, we draw upon two fundamental concepts from constructivist learning theory: experiential learning and situated cognition.

In the 1980s, David A. Kolb, an esteemed professor of organizational behavior in the United States, introduced the theory of experiential learning\cite{kolbExperientialLearningExperience2015}. Experiential learning accentuates the acquisition of knowledge and skills through firsthand experiences and practical engagement. Learners immerse themselves in authentic activities and contexts, where they encounter problems, endeavor to solve them, and accrue feedback and experiential wisdom.
Another pivotal concept is situated cognition theory, which was initially proposed by John Seely Brown and Allan Collins in the late 1980s. This theory underscores the cognitive processes of learners within specific contexts\cite{brownSituatedCognitionCulture1989}. Learners interact with their environment, observe, interpret, and comprehend information and events in relation to the context. Situated cognition theory posits that learning transpires through the connection and adjustment of novel knowledge with preexisting cognitive frameworks.

Drawing upon the concepts of experiential learning and situated cognition, the design of this study encompasses the following key aspects:

\subsubsection{Realistic Environment}
\begin{figure}
	\centering
	\includegraphics[width=\columnwidth]{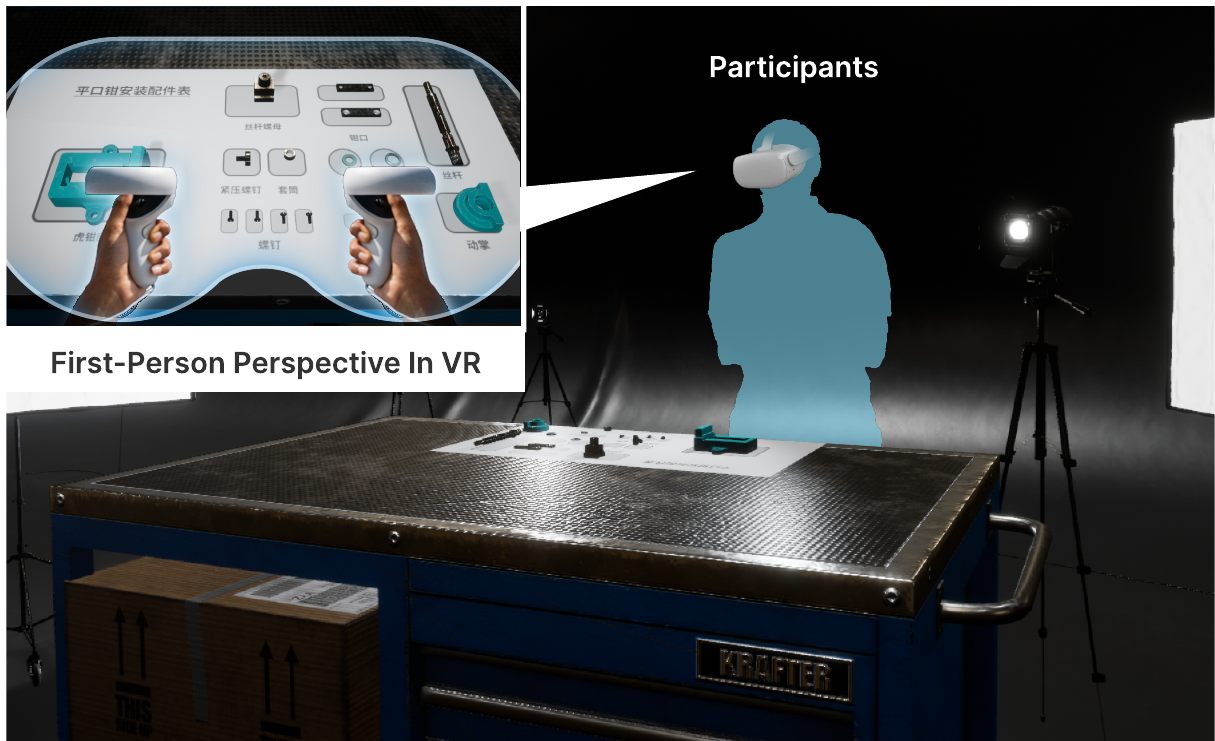}
	\caption{Virtual environment for training purposes, incorporating assembly models for training assembly tasks.}
	\label{fig:env}
\end{figure}

By employing advanced 3D scene modeling and high-fidelity environment rendering, the system furnishes a remarkably authentic virtual milieu that engrosses users and engenders focused learning by eliminating external distractions. Capitalizing on the realistic environment, the system further addresses assembly predicaments within the virtual domain, thereby imbuing virtual simulation training with an unparalleled sense of authenticity. Consequently, the virtual environment in this system emulates a conventional assembly training workstation, as depicted in Figure \ref{fig:env}

\subsubsection{Appropriate Level of Difficulty}
\begin{figure*}
	\centering
	\includegraphics[width=\textwidth]{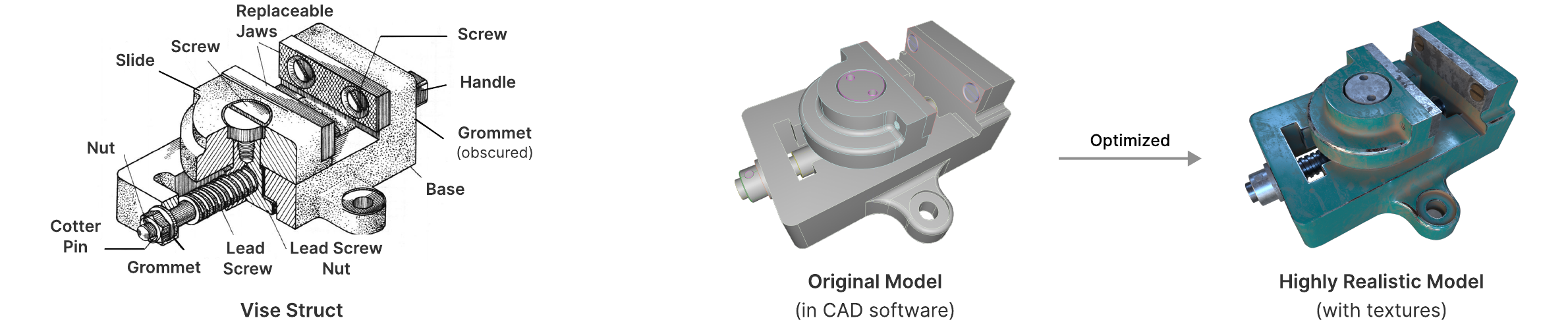}
	\caption{The left side shows the physical structure of the bench vise, and the right side shows the change from the original CAD model to a highly realistic model with optimized topology and applied material textures.}
	\label{fig:vise_spec}
\end{figure*}

Students who lack prior experience with virtual structural cognition training necessitate an adaptation period when employing the system. This necessitates ample guidance, user-friendly interfaces, and logical interactions. Moreover, the practical activities involved possess a relatively specialized nature, hence the difficulty level of these activities should be moderate. Setting the difficulty too high may instigate student anxiety and a lack of confidence, while setting it too low can breed tedium and disinterest. The system selects a bench vise as the training material due to its simple structure and appropriate assembly difficulty, rendering it an apt choice, as depicted in Figure \ref{fig:vise_spec}. This figure also illustrates the rendered outcomes of the optimized model within the scene.

%


\subsubsection{Prudent Guidance}
\begin{figure}
	\centering
	\includegraphics[width=\columnwidth]{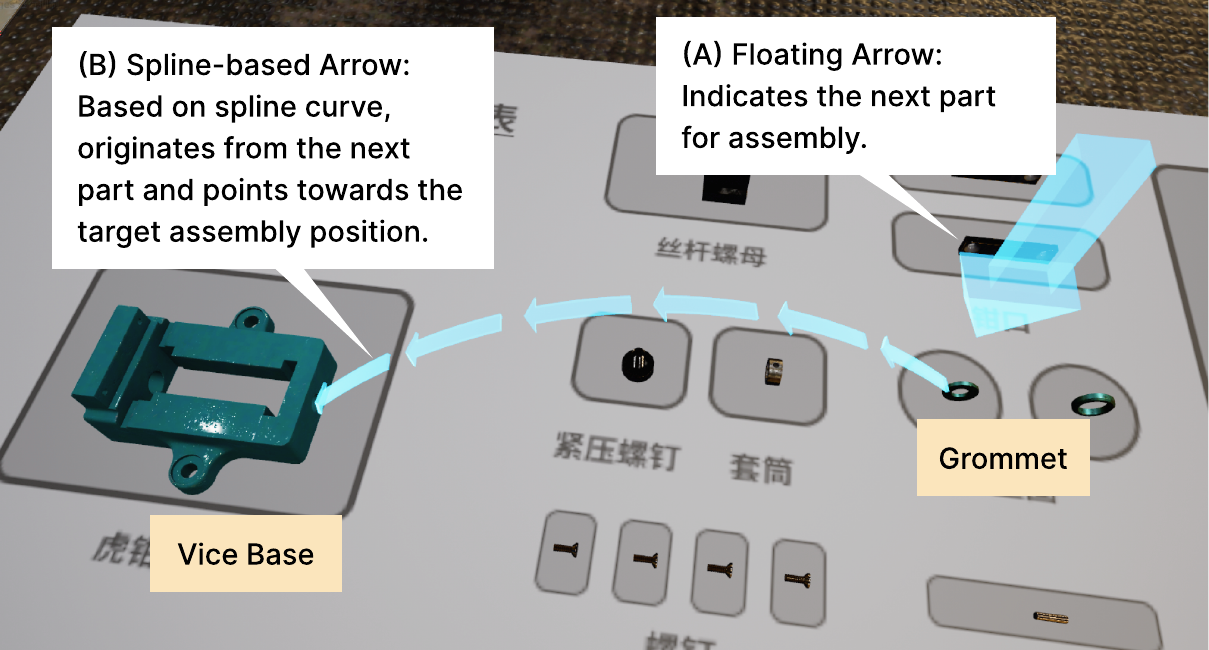}
	\caption{An exemplar of using two types of arrows (A.Floating arrow B.Spline-based arrow) in a specific scenario.}
	\label{fig:arrow_example}
\end{figure}

   Inadequate task guidance design within the system can precipitate various challenges for students, including uncertainty regarding navigation within the virtual environment or interaction with virtual elements. Such deficiencies not only impede the immersive practical intent but also significantly diminish the user experience, erode student confidence, and dampen enthusiasm for learning. Therefore, the design must strive to minimize cognitive load on users. Our solution entails the use of floating arrows and spline-based arrows within the scene, facilitating guidance on the assembly relationship between two components. Figure \ref{fig:arrow_example} presents a practical example of utilizing instructional arrows within the system.


\subsubsection{Natural Interaction}

   This form of interaction seeks to emulate the manipulation of objects in a physical environment, akin to the way they are handled with hands. It mandates that interactive objects be within the user's reachable range, such as utilizing a grab button on a controller to replicate the grasping action observed in a real-world physical environment when manipulating virtual objects. Direct interaction proves more intuitive and comprehensible for users, reducing their cognitive burden.

\subsection{Engine Selection}
The game engine plays a crucial role in performing graphics calculations and rendering 3D objects with stereoscopic effects on a 2D screen. In VR development, Unity and Unreal Engine 4 are the leading options.

Unreal Engine 4, developed by Epic Games, features Blueprint, a visual scripting system reducing programming complexities, aiding novice developers in quick feature implementation. Unreal Engine stands out in Physically Based Rendering (PBR) workflows and boasts a robust material editor, delivering impressive visual performance and realistic environments\cite{winVisualEffectsCreation2021}. Its comprehensive toolset and thriving community ecosystem have gained popularity in film animation and architecture.


Considering the requirement for demanding artistic expression and high realism in this system, Unreal Engine 4 has been chosen as the development tool. While Unity 3D is another viable option, Unreal Engine 4 aligns better with our goals for achieving realistic visuals and high-performance environments.


%
%
%

\subsection{Development Pipeline}
The study focuses on developing two structural cognition training systems using Unreal Engine 4 for VR and desktop platforms. The systems revolve around interacting with a machine vise as the primary object. They were deployed on Oculus Quest 2 VR headsets and Windows 10-based PCs for experiments, testing, and comparative evaluations to assess their efficacy.

It's important to note that VR devices, like the Oculus Quest 2, have limited computational power similar to smartphones\cite{rostamiMetaverseImplementingAdvanced2022}. Therefore, optimization measures are crucial to achieve real-time rendering of high-quality visuals and ensure smooth system performance. Optimization can be approached from two perspectives: model optimization , UV optimization and rendering optimization.

\subsubsection{Model Optimization}
Model optimization involves topology reduction and UV processing. Optimizing the model's topology is necessary to reduce data size and improve program execution speed\cite{yuanSimplifiedTessellatedMesh2016}. Autodesk$^\circledR$ Maya$^\circledR$ (version used: 2019) was employed to optimize the model while ensuring minimal loss of geometric form and detailed structure. The Quadrangulate function in Maya's Marking Menu was utilized to convert the model into a quadrilateral structure. For more intricate quadrilateral models, the QuadRemesher plugin facilitated automated reduction or manual topology adjustment.

\subsubsection{UV Optimization}
UV texture mapping coordinates pertains to the process of projecting each point on a 3D model onto a 2D plane image\cite{yangDeepFaceVideo2023}. Traditional UV mapping corresponds to a single UV texture within the UV range (0-1), with the entire mesh texture possessing a single resolution. However, with the growing demand for visual effects, a single quadrant UV is no longer adequate to meet the requirements. UDIM technology offers a solution by utilizing a tiling system where each block space forms its own texture within the overall UDIM array. UDIM technology allows multiple models to share the same UV system, conserving computational and storage resources.



\subsubsection{Rendering Optimization}
Unreal Engine 4 uses the Deferred Renderer by default, but for VR projects, Forward Rendering provides better performance. Forward Rendering offers a faster baseline rendering pipeline and superior anti-aliasing effects\cite{hendrickxRealisticRenderingVirtual2011}. Multi-Sample Anti-Aliasing (MSAA) is preferred in VR due to sub-pixel movement caused by head tracking.




\begin{figure*}
	\centering
	\includegraphics[width=\textwidth]{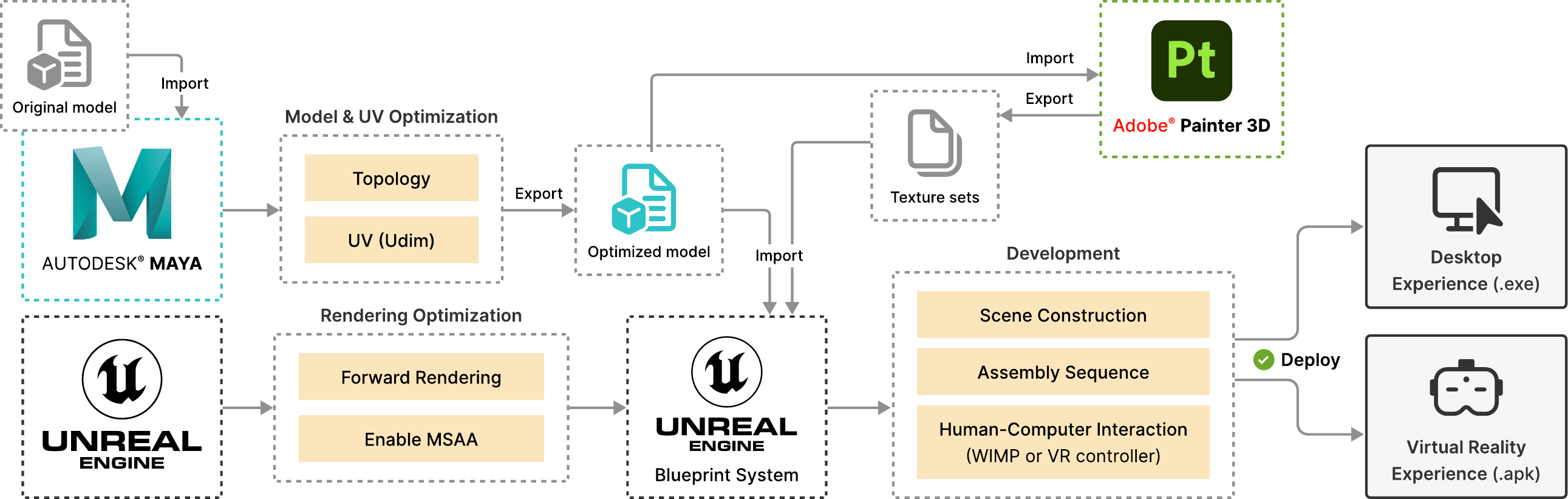}
	\caption{The development pipeline of this system.}
	\label{fig:pipeline}
\end{figure*}

After completing the optimization settings, the specific interaction logic can be developed within the engine. The development pipeline is depicted in Figure \ref{fig:pipeline}. It is worth noting that both configurations share a significant portion of software modules. Consequently, for both configurations, the generation of task-related visual elements (labels, colors, animations, etc.) and auditory information is handled in the same manner.

\section{Experimental study}
Our experimental investigation took place at the User Experience Research Laboratory at South China University of Technology (SCUT). Adhering to Mayer's principles of instructional experience design\cite{mayerComputerGamesLearning2014}, this study aimed to compare the immersive training experience in virtual reality (VR) with two conventional teaching methods: desktop-based assembly training and physical assembly training. The primary objective was to examine potential variations in learning outcomes across the three training methods. The secondary objective was to delve into the correlation between presence in the virtual environment and the training's learning efficacy.

\subsection{Sample Size}
To determine the sample size for our experiment, we utilized the G*Power software. A minimum sample size of 51 was determined based on calculations. Recruitment announcements were distributed among engineering graphics courses, resulting in 53 voluntary registrations online. Of the participants, 17 were female (32.08\%), and 36 were male (67.92\%), with ages ranging from 19 to 26. None had prior experience in bench vise assembly, which aligns with the training's focus on novices. Random assignment allocated individuals to three groups: VR (24 students), desktop (22 students), and physical assembly (7 students). To investigate the relationship between presence in the virtual environment and training efficacy, the sample size for the VR and desktop groups was intentionally increased.

\subsection{Procedure}

For each participant, the following steps were conducted:

\subsubsection{Pretest}
Participants were briefed on the experiment's expected duration (about 20 minutes) and provided a detailed procedure overview.  It commenced with a pretest consisting of eight single-choice questions and one image-based fill-in-the-blank question.  Single-choice questions gauged assembly structure comprehension, while the fill-in-the-blank assessed key component name familiarity.  After the pretest, experts from the research team provided a systematic explanation of the composition and assembly process of the bench vise to address any weaknesses identified during the pretest. The specific pretest questions can be found in Appendix 1.


\subsubsection{Training Practice}
Two days later, the three groups engaged in their respective learning experiences: using the VR system, the desktop system, or physically assembling the bench vise. For the VR group, an additional 5 minutes of guidance and free practice time focused on object selection, movement, and placement\cite{serranovergelComparativeEvaluationVirtual2020}. The VR and desktop groups performed the assembly tasks within their respective virtual environments, while the physical assembly group utilized a physical model of the bench vise provided by the institute to complete the task.

\subsubsection{Evaluation}

After completing the training tasks, the desktop and VR groups were asked to fill out a questionnaire. Appendix 2 lists the questions raised in this survey questionnaire. These questions are categorized into the five factors proposed by Witmer cite for measuring the sense of presence in a virtual environment\cite{witmerMeasuringPresenceVirtual1998}: Sensory Factors (SF), Control Factors (CF), Distraction Factors (DF), Realism Factors (RF), and Ergonomic Factors (EF). The SF evaluates the support for participants' perceptual experience in aspects like sensory stimuli, environment richness, information richness, multimodal presentation, and cross-modal consistency; the CF examines the smoothness of interactive operations by looking into issues like control latency and conformity of system response to expectation; the DF checks the efficacy of external distraction isolation to prevent disruption of immersion; the RF examines the verisimilitude and congruity of virtual scenes to elicit the sensation of physical presence within the environment; the EF surveys the human-computer interaction friendliness in the virtual environment to match participants' operational habits. This questionnaire was adapted from Witmer's Presence Questionnaire (PQ) for measuring presence in virtual environments, hence the physical assembling group participants did not need to complete it.

\subsubsection{Post-test}

A knowledge test was conducted five days after the assembly training to assess the training's long-term effects\cite{mayerComputerGamesLearning2014}.  Immediate testing right after the training would primarily assess short-term memory and might not reflect the long-term efficacy of the training. The post-test comprised 20 multiple-choice questions and 1 image-based fill-in-the-blank question. The post-test questions differed from the pretest to prevent potential bias resulting from participants' familiarity. All post-test questions were new, and the format changed from single-choice to multiple-choice, increasing the difficulty level.  Specific post-test questions can be found in the Appendix 3.


\section{Results and Discussion}

This section presents the statistical analysis outcomes obtained from the assembly training experiment. The raw data and analysis are available in Appendix 4. Two-sided significance testing at a 0.05 level was conducted for all analyses herein using Jamovi\cite{jamovi_2022,R_2021,car_package} and MATLAB\cite{MATLAB_2022b} software. The analysis aims to compare the effects of three learning methods—physical, desktop, and VR—on learning efficacy. Thus, analysis of variance was utilized to contrast post-test score differences between the 3 subject groups. A confounding factor exists in the form of prior knowledge levels before the experiment. Hence, pre-test scores were employed as a covariate in a one-way ANCOVA of post-test scores. Additionally, the analysis seeks to determine if significant differences in presence exist between the two virtual training systems, and whether presence correlates with learning efficacy. By examining this association, the extent of presence's influence on learning efficacy can be further elucidated.

\subsection{Learning Efficacy (One-way ANCOVA)  }

\subsubsection{Assumption Checks}

%

\begin{table*}
	\mylineheight
	\caption{Description and Distribution of Pre-test and Post-test}
	\label{tab:des_test}
	\centering
	\begin{tabular}{c   || c      c     S[table-format=2.3,table-column-width=1cm]@{ ±}        S[table-format=2.3,table-column-width=1cm]      c  |   S[table-format=2.3]   S[table-format=2.3]    S[table-format=2.3]     c  |  c   c }
		\mytoprule
		& \textbf{Group} & \textbf{N} & \multicolumn{2}{c}{\textbf{Mean ± SD.}} & \textbf{Medium} & \textbf{IQR} & \textbf{Q1-1.5.IQR} & \textbf{Q3+1.5IQR} & \textbf{Outliner} & \textbf{Shapiro-Wilk's W} & \textbf{\textit{p}}  \\
		\mymidrule
		\multirow{3}*{\textbf{pre}} & Physical   & 7     & 26.786 & 7.319    & 30.000       &7.500   &11.250     &41.250  & 0     & 0.937     & 0.609  \\
		& Desktop    & 22    & 28.864 & 8.116    & 30.000       &14.125  &1.438      &57.938  & 0     & 0.921     & 0.079   \\
		& VR         & 24    & 26.458 & 10.186   & 26.250       &13.125  &0.000    &51.563  & 0     & 0.945     & 0.210   \\
		\hline
		\multirow{3}*{\textbf{post}}    & Physical  & 7      & 61.714 & 5.707    & 60.000   &9.000  &43.500     &79.500  & 0     & 0.892     & 0.288  \\
		& Desktop   & 22     & 64.364 & 6.807    & 65.000   &8.000   &48.000     &80.000  & 0     & 0.950     & 0.318   \\
		& VR        & 24     & 69.417 & 5.090    & 70.000   &6.000   &57.000     &81.000  & 1     & 0.935     & 0.127   \\
		\mybottomrule
	\end{tabular}
\end{table*}

Using pre-test scores as a covariate in one-way ANCOVA necessitates that no significant pre-test score differences exist across groups. Thus, an analysis of variance on pre-test scores for the three groups was first conducted. The pre-test scores are certainly independent and continuous. Moreover, the pre-test data fulfills normality (Table \ref{tab:des_test}, \textit{p}$>$0.05 in all groups per Shapiro-Wilk test) and homogeneity of variance (non-significant Levene’s test, F=1.380, \textit{p}=0.261) assumptions, with no significant outliers (Table \ref{tab:des_test}). This satisfies ANOVA assumptions. Testing showed no significant pre-test score differences between groups (F=0.431, \textit{p}=0.652, $\eta^{2}$=0.017).

\begin{figure}
	\centering
	\subfloat[]{\includegraphics[height=1.34in]{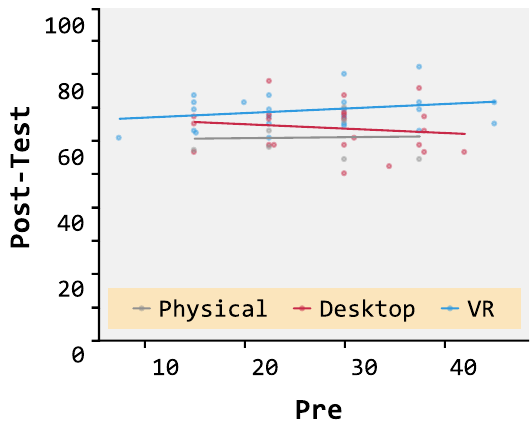}}
	\hfill
	\subfloat[]{\includegraphics[height=1.34in]{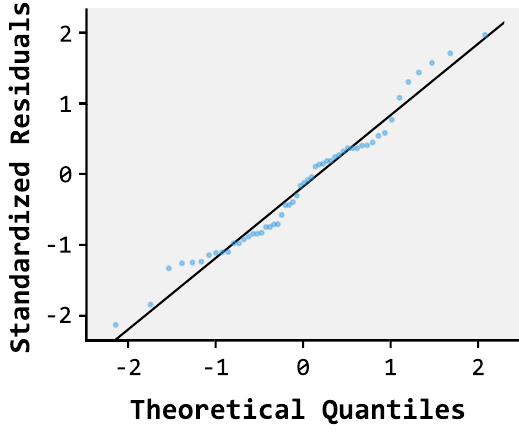}}
	\hfill
	\subfloat[]{\includegraphics[height=1.34in]{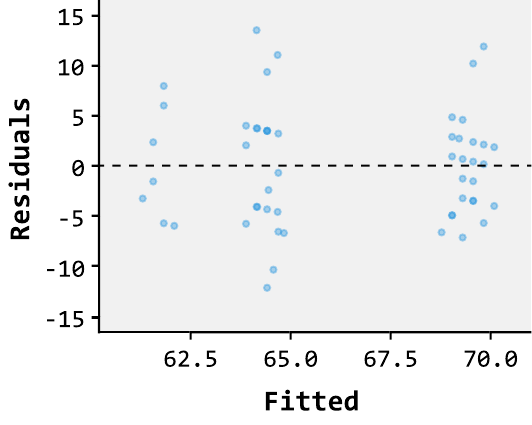}}
	\hfill
	\subfloat[]{\includegraphics[height=1.34in]{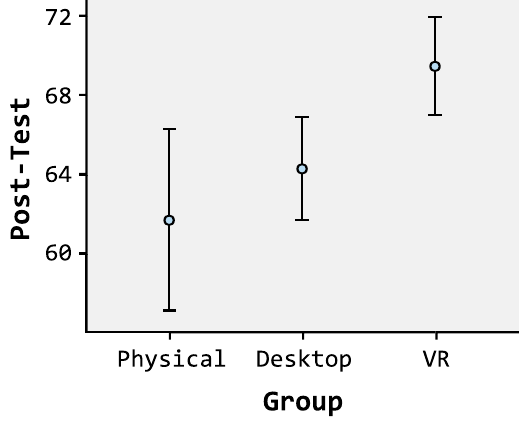}}
	\caption{Validation of ANCOVA assumptions and data description: (a) Scatter plot displaying linear relationship between pre-test and post-test; (b) Q-Q plot showing normality of residuals; (c) Residuals scatter plot; (d) Marginal means plot for statistical description..}
	\label{fig:ancova_valid}
\end{figure}


Clearly, post-test data is also independent and continuous, meeting normality (Table \ref{tab:des_test}, \textit{p}$>$0.05 in all groups per Shapiro-Wilk test) and homogeneity of variance (non-significant Levene’s test, F=1.795, \textit{p}=0.177) assumptions, despite one VR outlier (Table \ref{tab:des_test}). However, experts considered this outlier to be within a reasonable range, hence it was retained. Furthermore, testing showed normally distributed post-test data residuals (Shapiro-Wilk Statistic=0.973, \textit{p}=0.259) exhibiting homoscedasticity (Figure \ref{fig:ancova_valid}(c), random, even scatter around y=0 in fitted residual plot). The covariate and dependent variable also meet homogeneity of regression slopes (Non-significant interaction between group and pre-test, \textit{p}=0.471). In summary, the post-test data meets ANCOVA prerequisites.

\subsubsection{Descriptive Statistics }


\begin{figure}
	\centering
	\subfloat[]{\includegraphics[height=1.24in]{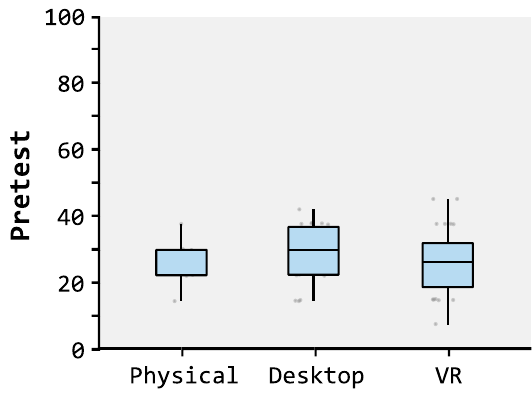}}
	\hfill
	\subfloat[]{\includegraphics[height=1.24in]{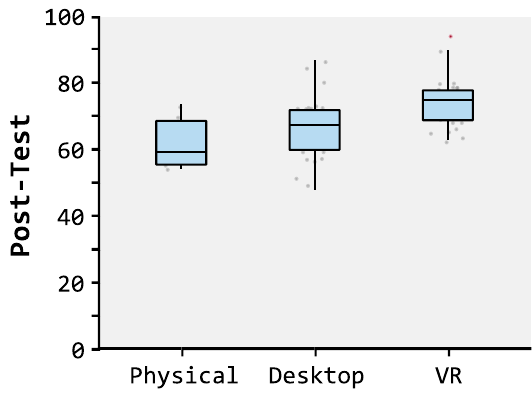}}
	\caption{Boxplots of Pre-test and Post-test.}
	\label{fig:box}
\end{figure}

Table \ref{tab:des_test} displays pre-test scores of 26.786 ± 7.319, 28.864 ± 8.116, and 26.458 ± 10.186 for the three groups, respectively. Respective post-test scores were 61.714 ± 5.707, 64.364 ± 6.807, and 69.417 ± 5.090. Boxplots of pre-test and post-test scores are shown in Figure \ref{fig:box}. Estimated marginal post-test score means of 61.739 (95\% CI: 57.186$\sim$66.293), 64.316 (95\% CI: 61.736$\sim$66.896), and 69.453 (95\% CI: 66.987$\sim$71.919) for the physical, desktop, and VR groups. It can be seen that there are substantial differences between the raw mean post-test scores of the three groups and between the estimated marginal mean post-test scores of the three groups. Figure \ref{fig:ancova_valid}(d) (estimated marginal means plot) also depicts the post-test scores, with VR being highest, desktop second, and physical lowest. However, statistical inference is required to determine if these differences are significant.

\subsubsection{Statistical Inference}

\begin{table}
	\caption{ANCOVA\label{tab:ancova}}
	\mylineheight
	\centering
	\begin{tabular}{l||
			S[table-format=4.3]
			S[table-format=2.0]
			S[table-format=1.3]
			S[table-format=1.3]
			S[table-format=1.3]}
		\mytoprule
		& \textbf{Sum of squares} &\textbf{df}   & \textbf{\textit{F}} &\textbf{\textit{p}} &    \textbf{$\eta^{2}$}                                     \\
		\mymidrule
		\textbf{group}      & 467.605       & 2      & 6.512    & 0.003     & 0.210    \\
		\textbf{pre}        & 5.004         & 1      & 0.139    & 0.711     & 0.002 \\
		\textbf{Residuals}  & 1759.349      & 49 \\
		\mybottomrule
	\end{tabular}
\end{table}

A one-way ANCOVA was conducted to examine the effects of different learning methods on learning outcomes, with the pretest as covariate. The results showed a significant main effect of learning method on posttest scores (Table \ref{tab:ancova}, F = 6.512, \textit{p} = 0.003, $\eta^{2}$=0.210), indicating that the group category accounted for about 21\% of total variance. This suggests that the inter-group variation has considerable explanatory power for the overall variation.


Therefore, it can be concluded that, given no significant differences in prior knowledge levels before the commencement of the experiment, varying methods of assembly training can engender significant between-group effects on performance.

\subsection{Presence}

Presence aspects were further scrutinized. Reliability and validity analyses ensured adequate reliability of the presence scale. Reliability analysis evaluated scale stability, consistency across conditions, and construct measurement accuracy. With reliability and validity confirmed, VR and desktop systems were compared on presence. Finally, the presence-efficacy correlation was explored.


\subsubsection{Reliability Analysis}

\begin{table}
	\caption{Significance of Factor Loadings and Modification Indices for Item-Factor Associations under Five Factors}
	\label{tab:srs}
	\mylineheight
	\centering
	\begin{tabular}{c||
			S[table-format=1.3] |
			S[table-format=1.3]
			S[table-format=1.3]
			S[table-format=1.3]
			S[table-format=1.3]
			S[table-format=1.3] }
		\mytoprule
		&	\textbf{\textit{p}}& \textbf{CF} &\textbf{SF}   & \textbf{DF} & \textbf{RF} &  \textbf{EF}    \\
		&	{Loading}& {Q01-15} &{Q16-26 } & {Q27-33} &{Q34-39} & {Q40-44}    \\
		\mymidrule

		Q01	&	—	&   —	&	0.000	&	0.333	&	0.005	&	0.694	\\
		Q02	&	0.014	&   —	&	0.673	&	8.302	&	1.002	&	3.934	\\
		Q03	&	0.017	&   —	&	1.404	&	0.004	&	1.083	&	0.004	\\
		Q05	&	0.044	&   —	&	0.304	&	0.517	&	0.648	&	6.349	\\
		Q06	& 	\textless $\mathrm{.001} $ &   —	&	0.228	&	0.019	&	0.182	&	1.854	\\
		Q07	&	0.018	&   —	&	0.702	&	0.238	&	0.305	&	0.189	\\
		Q08	&	\textless $\mathrm{.001} $	&   —	&	2.159	&	0.024	&	0.195	&	0.268	\\
		Q09	&	0.014	&   —	&	2.338	&	0.229	&	0.169	&	1.319	\\
		Q10	&	\textless $\mathrm{.001} $	&   —	&	1.970	&	0.027	&	2.876	&	0.727	\\
		Q11	&	0.003	&   —	&	0.034	&	0.149	&	0.723	&	2.554	\\
		Q12	&	\textless $\mathrm{.001} $	&   —	&	0.032	&	0.217	&	0.501	&	1.548	\\
		Q13	&	\textless $\mathrm{.001} $	&   —	&	0.207	&	0.188	&	0.226	&	0.656	\\
		Q14	&	0.005	&   —	&	0.112	&	0.066	&	0.015	&	1.325	\\
		Q15	&	0.026	&   —	&	0.279	&	0.217	&	0.064	&	0.459	\\
		\hline
		Q16	&	—	&	6.513	&	—	&	6.486	&	1.599	&	5.594	\\
		Q17	&	0.004	&	0.783	&	—	&	6.390	&	3.051	&	0.029	\\
		Q18	&	0.002	&	1.073	&	—	&	1.203	&	1.316	&	1.400	\\
		Q19	&	0.023	&	0.243	&	—	&	2.359	&	1.806	&	0.930	\\
		Q20	&	0.002	&	0.000	&	—	&	2.013	&	9.006	&	0.669	\\
		Q21	&	0.007	&	1.875	&	—	&	3.329	&	0.644	&	0.459	\\
		Q22	&	0.010	&	1.454	&	—	&	0.757	&	0.629	&	0.100	\\
		Q23	&	0.002	&	0.073	&	—	&	4.592	&	0.133	&	0.597	\\
		Q24	&	0.002	&	4.643	&	—	&	0.092	&	6.772	&	\underline{14.631}	\\
		Q25	&	0.001	&	0.272	&	—	&	1.610	&	2.182	&	2.127	\\
		Q26	&	0.002	&	0.200	&	—	&	0.244	&	3.598	&	0.996	\\
		\hline
		Q27	&	—	&	7.629	&	1.283	&	—	&	0.343	&	1.214	\\
		Q30	&	\textless $\mathrm{.001} $	&	2.051	&	0.191	&	—	&	2.898	&	3.373	\\
		Q31	&	\textless $\mathrm{.001} $	&	0.008	&	0.594	&	—	&	0.347	&	1.445	\\
		Q32	&	\textless $\mathrm{.001} $	&	3.485	&	4.030	&	—	&	1.420	&	1.939	\\
		Q33	&	\textless $\mathrm{.001} $	&	6.198	&	0.282	&	—	&	6.225	&	5.759	\\
		\hline
		Q34	&	—	&		1.318	&	0.717	&	2.786	&	—	&	2.226	\\
		Q35	&	0.014	&		0.133	&	3.906	&	2.199	&	—	&	0.015	\\
		Q36	&	0.013	&		4.593	&	0.765	&	9.361	&	—	&	6.777	\\
		Q37	&	0.008	&		2.623	&	0.018	&	0.081	&	—	&	1.217	\\
		Q39	&	0.007	&		0.275	&	0.008	&	1.041	&	—	&	0.931	\\
		\hline
		Q40	&	—	&	0.603	&	1.165	&	1.568	&	0.131	&	—	\\
		Q41	&	0.009	&	2.692	&	0.084	&	0.069	&	0.154	&	—	\\
		Q42	&	0.089	&	1.903	&	0.320	&	0.000	&	0.151	&	—	\\
		Q43 &	0.078	&	0.149	&	0.345	&	2.623	&	0.707	&	—	\\
		Q44	&	0.100	&	0.511	&	0.603	&	7.013	&	1.560	&	—	\\
		\mybottomrule
	\end{tabular}
\end{table}

Internal consistency was assessed via Cronbach’s $\alpha$. Results were 0.860, 0.881, 0.755, 0.674, and 0.777 for CF, SF, DF, RF, and EF, indicating very good consistency for CF and SF, relatively good consistency for DF and EF, and acceptable consistency for RF. Thus, overall data reliability is good, enabling further analysis.

\subsubsection{Validity Analysis}

Regarding validity, the scale (Appendix 2) was obtained by further specializing the scenario based on the "five factors for measuring presence in virtual environments" summarized from previous research. The design of the scale was approved by experts in the research group, so content validity can be ensured. Since existing dimensions capturing presence influences are sufficiently comprehensive, exploratory factor analysis was unnecessary. Only confirmatory factor analysis was required to evaluate factor-item alignment with expectations.

Overall, the vast majority of items exhibited significance at the 0.05 level (\textit{p}$<$0.05) with standardized factor loadings above 0.7, indicating satisfactory overall factor-item correspondence and aggregation validity. Items with \textit{p}$>$0.05 (Q4, Q28, Q29, Q38) were removed and analysis redone. Table \ref{tab:srs} presents the modification indices (MI) for factor-item associations. Except for Q24, the correspondence was satisfactory for other items. Though Q24 belonged to the SF dimension, it had a high MI in the EF dimension as Q24 examined whether users could view objects in the virtual environment from multiple angles, which is highly device-dependent. Virtual reality devices can achieve this easily while desktop applications require coordinated mouse and keyboard operations for observing a specific angle through multiple steps. Thus, the high Q24 - EF association is understandable. However, considering Witmer included active search in the SF dimension and Q24 examines such capability, Q24 was retained in the SF dimension.


\begin{subequations}\label{equ:avencr}
	\begin{align}
		AVE&=\frac{\Sigma {\lambda }^{2}}{\left[\sum {\lambda }^{2}+\Sigma \left(1-{\lambda }^{2}\right)\right]}  \label{asda} \\
		CR&=\frac{{\left(\Sigma \lambda \right)}^{2}}{\left[{\left(\Sigma \lambda \right)}^{2}+\sum \left(1-{\lambda }^{2}\right)\right]}  \label{asdasssssssssd}
	\end{align}
\end{subequations}

Regarding the convergent validity, calculate the AVE (Average Variance Extracted) and CR (Composite Reliability) values for the five factors using their respective equations (Equation \ref{equ:avencr}, $\lambda$ represents the factor loadings for each item within their respective factors), as shown in Table \ref{tab:rmatrix}. Not all AVE values were greater than 0.5, but CR values exceeded 0.7. Thus, AVE of 0.36$\sim$0.5 is acceptable \cite{lamImpactCompetitivenessSalespeople2012,fornellEvaluatingStructuralEquation1981} , indicating excellent convergence.

For discriminant validity, the correlation matrix (Table \ref{tab:rmatrix}) shows CF Sqrt(AVE) of 0.602 exceeded correlations with other factors (Figure \ref{fig:heatmap}, max 0.395). SF Sqrt(AVE) of 0.687 exceeded correlations with other factors (max 0.591). Similarly, Sqrt(AVE) values for DF, RF, and EF exceeded correlations. This demonstrates good discriminant validity. Additionally, no $>$50\% variance explained by any factor indicated no common method bias.

\begin{table}
	\caption{Factor Statistics  }
	\label{tab:rmatrix}
	\mylineheight
	\centering
	\begin{tabular}
		{ c ||S[table-format=2.3]
			S[table-format=2.3]
			S[table-format=2.3]
			S[table-format=2.3]
			S[table-format=2.3]}
		\mytoprule
		&  \textbf{CF} & \textbf{SF} & \textbf{DF} & \textbf{RF} & \textbf{EF}                                               \\
		\mymidrule

		\textbf{AVE}              &0.362  &0.472  &0.554  &0.367  &0.440\\
		\textbf{CR}               &0.863  &0.904  &0.861  &0.739  &0.783\\
		\textbf{Sqrt of AVE}        &0.602  &0.687  &0.744  &0.606  &0.663 \\

		\mybottomrule
	\end{tabular}
\end{table}


\subsubsection{Data Description}


\begin{table*}
	\caption{Description of Scale Data \& Mann-Whitney U Test}
	\label{tab:des_scale}
	\centering
	\mylineheight
	\begin{tabularx}{0.976\textwidth}{
			>{\centering\arraybackslash}p{0.5cm}  ||
			>{\centering\arraybackslash}p{0.8cm}
			>{\centering\arraybackslash}p{0.4cm}
			>{\centering\arraybackslash}p{0.8cm}
			>{\centering\arraybackslash}p{0.8cm}
			>{\centering\arraybackslash}p{0.9cm}
			>{\centering\arraybackslash}p{0.9cm}
			>{\centering\arraybackslash}p{0.9cm}
			>{\centering\arraybackslash}p{0.9cm}
			>{\centering\arraybackslash}p{0.55cm}
			>{\centering\arraybackslash}p{0.75cm}
			>{\centering\arraybackslash}p{1.2cm}
			>{\centering\arraybackslash}p{1.4cm}
			>{\centering\arraybackslash}p{1.4cm} }
		\mytoprule
		\vspace{0.1em}
		& \multirow{2}{*}{\textbf{group}}
		& \multirow{2}{*}{\textbf{N}}
		& \multicolumn{2}{c}{\multirow{2}*{\textbf{Mean ± SD.}}}
		& \multirow{2}{*}{\textbf{Med.}}
		& \multicolumn{2}{c}{\textbf{Shapiro-Wilk}}
		& \multirow{2}{*}{\textbf{U}}
		& \multirow{2}{*}{\textbf{\textit{p}}}
		& \multirow{2}{*}{\textbf{Mean d.}}
		& \multirow{2}{*}{\textbf{Corelation}}
		& \multicolumn{2}{c}{\textbf{95\% Confidence interval}}   \\
		\cline{7-8}
		\cline{13-14}

		\vspace{0.1em}
		&
		&
		& \multicolumn{2}{c}{}
		&
		& \textbf{W}
		& \textbf{\textit{p}}
		&
		&
		&
		&
		& \textbf{Lower}
		& \textbf{Upper} \\
		\mymidrule
	\end{tabularx}

	\begin{tabularx}{0.976\textwidth}{
			>{\centering\arraybackslash}p{0.5cm}  ||
			>{\centering\arraybackslash}p{0.8cm}
			S[table-format=2.0,table-column-width=0.4cm]
			S[table-format=1.3,table-column-width=0.7cm]@{±}
			S[table-format=1.3,table-column-width=0.7cm]
			S[table-format=1.3,table-column-width=0.8cm]
			S[table-format=1.3,table-column-width=0.8cm]
			S[table-format=1.3,table-column-width=0.8cm]
			c
			c
			c
			c
			c
			c }

		\multirow{2}*{\textbf{CF}}
		&	Desktop	&	21	    &	6.184	&	0.570	&	6.286	&	0.888	&	0.021
		&\multicolumn{1}{c}{\multirow{2}{*}{\begin{tabular}{@{}S[table-format=3.3,table-column-width=0.9cm]@{}}192.500\end{tabular}}}
		&\multicolumn{1}{c}{\multirow{2}{*}{\begin{tabular}{@{}S[table-format=1.3,table-column-width=0.6cm]@{}}0.657\end{tabular}}}
		&\multicolumn{1}{c}{\multirow{2}{*}{\begin{tabular}{@{}S[table-format=1.3,table-column-width=0.9cm]@{}}0.071\end{tabular}}}
		&\multicolumn{1}{c}{\multirow{2}{*}{\begin{tabular}{@{}S[table-format=1.3,table-column-width=1.2cm]@{}}0.083\end{tabular}}}
		&\multicolumn{1}{c}{\multirow{2}{*}{\begin{tabular}{@{}S[table-format=1.3,table-column-width=1.4cm]@{}}-0.286\end{tabular}}}
		&\multicolumn{1}{c}{\multirow{2}{*}{\begin{tabular}{@{}S[table-format=1.3,table-column-width=1.4cm]@{}}0.500\end{tabular}}}
		\\
		&	VR	&	20	        &	6.100	&	0.573	&	6.071	&	0.933	&	0.177	\\

		\multirow{2}*{\textbf{SF}}
		&	Desktop	&	21	    &	5.281	&	0.942	&	5.455	&	0.933	&	0.155
		&\multicolumn{1}{c}{\multirow{2}{*}{\begin{tabular}{@{}S[table-format=3.3,table-column-width=0.9cm]@{}}103.000\end{tabular}}}
		&\multicolumn{1}{c}{\multirow{2}{*}{\begin{tabular}{@{}S[table-format=1.3,table-column-width=0.6cm]@{}}0.005\end{tabular}}}
		&\multicolumn{1}{c}{\multirow{2}{*}{\begin{tabular}{@{}S[table-format=1.3,table-column-width=0.9cm]@{}}-0.636\end{tabular}}}
		&\multicolumn{1}{c}{\multirow{2}{*}{\begin{tabular}{@{}S[table-format=1.3,table-column-width=1.2cm]@{}}0.510\end{tabular}}}
		&\multicolumn{1}{c}{\multirow{2}{*}{\begin{tabular}{@{}S[table-format=1.3,table-column-width=1.4cm]@{}}-1.182\end{tabular}}}
		&\multicolumn{1}{c}{\multirow{2}{*}{\begin{tabular}{@{}S[table-format=1.3,table-column-width=1.4cm]@{}}-0.273\end{tabular}}}
		\\
		&	VR	&	20	        &	6.050	&	0.596	&	5.909	&	0.923	&	0.115	\\

		\multirow{2}*{\textbf{DF}}
		&	Desktop	&	21	    &	6.010	&	0.835	&	6.200	&	0.880	&	0.015
		&\multicolumn{1}{c}{\multirow{2}{*}{\begin{tabular}{@{}S[table-format=3.3,table-column-width=0.9cm]@{}}181.500\end{tabular}}}
		&\multicolumn{1}{c}{\multirow{2}{*}{\begin{tabular}{@{}S[table-format=1.3,table-column-width=0.6cm]@{}}0.462\end{tabular}}}
		&\multicolumn{1}{c}{\multirow{2}{*}{\begin{tabular}{@{}S[table-format=1.3,table-column-width=0.9cm]@{}}-0.200\end{tabular}}}
		&\multicolumn{1}{c}{\multirow{2}{*}{\begin{tabular}{@{}S[table-format=1.3,table-column-width=1.2cm]@{}}0.136\end{tabular}}}
		&\multicolumn{1}{c}{\multirow{2}{*}{\begin{tabular}{@{}S[table-format=1.3,table-column-width=1.4cm]@{}}-0.600\end{tabular}}}
		&\multicolumn{1}{c}{\multirow{2}{*}{\begin{tabular}{@{}S[table-format=1.3,table-column-width=1.4cm]@{}}0.200\end{tabular}}}
		\\
		&	VR	&	20	        &	6.180	&	0.731	&	6.300	&	0.904	&	0.050	\\

		\multirow{2}*{\textbf{RF}}
		&	Desktop	&	21	    &	4.448	&	0.953	&	4.400	&	0.958	&	0.481
		&\multicolumn{1}{c}{\multirow{2}{*}{\begin{tabular}{@{}S[table-format=3.3,table-column-width=0.9cm]@{}}93.000\end{tabular}}}
		&\multicolumn{1}{c}{\multirow{2}{*}{\begin{tabular}{@{}S[table-format=1.3,table-column-width=0.6cm]@{}}0.002\end{tabular}}}
		&\multicolumn{1}{c}{\multirow{2}{*}{\begin{tabular}{@{}S[table-format=1.3,table-column-width=0.9cm]@{}}-1.000\end{tabular}}}
		&\multicolumn{1}{c}{\multirow{2}{*}{\begin{tabular}{@{}S[table-format=1.3,table-column-width=1.2cm]@{}}0.557\end{tabular}}}
		&\multicolumn{1}{c}{\multirow{2}{*}{\begin{tabular}{@{}S[table-format=1.3,table-column-width=1.4cm]@{}}-1.400\end{tabular}}}
		&\multicolumn{1}{c}{\multirow{2}{*}{\begin{tabular}{@{}S[table-format=1.3,table-column-width=1.4cm]@{}}-0.400\end{tabular}}}
		\\
		&	VR	&	20	        &	5.360	&	0.860	&	5.200	&	0.962	&	0.582	\\

		\multirow{2}*{\textbf{EF}}
		&	Desktop	&	21	    &	6.619	&	0.473	&	6.800	&	0.792	&	\textless $\mathrm{.001} $
		&\multicolumn{1}{c}{\multirow{2}{*}{\begin{tabular}{@{}S[table-format=3.3,table-column-width=0.9cm]@{}}138.000\end{tabular}}}
		&\multicolumn{1}{c}{\multirow{2}{*}{\begin{tabular}{@{}S[table-format=1.3,table-column-width=0.6cm]@{}}0.055\end{tabular}}}
		&\multicolumn{1}{c}{\multirow{2}{*}{\begin{tabular}{@{}S[table-format=1.3,table-column-width=0.9cm]@{}}0.200\end{tabular}}}
		&\multicolumn{1}{c}{\multirow{2}{*}{\begin{tabular}{@{}S[table-format=1.3,table-column-width=1.2cm]@{}}0.343\end{tabular}}}
		&\multicolumn{1}{c}{\multirow{2}{*}{\begin{tabular}{@{}S[table-format=1.3,table-column-width=1.4cm]@{}}0.000\end{tabular}}}
		&\multicolumn{1}{c}{\multirow{2}{*}{\begin{tabular}{@{}S[table-format=1.3,table-column-width=1.4cm]@{}}0.800\end{tabular}}}
		\\
		&	VR	&	20	        &	6.210	&	0.718	&	6.300	&	0.880	&	0.018	\\
		\mybottomrule
	\end{tabularx}
\end{table*}

Examining presence differences between desktop and VR required determining normality conformity. Multiple datasets violated normality (Table \ref{tab:des_scale}), necessitating Mann-Whitney U testing.

\subsubsection{Non-parametric Testing}

The Mann-Whitney U test (Table \ref{tab:des_scale}) showed: desktop and VR groups exhibited significant differences in the SF and RF, but no significant differences in CF, DF and EF. This can be explained as follows:

The SF measures the richness of sensory experience including visual, auditory and other information. Compared to the desktop-based system, the VR system provided a more realistic 3D environment and immersive multisensory feedback, which directly enhanced the user's environmental perception. The RF examines the realism and consistency of virtual scenes. The VR system greatly improved the scene realism so that users felt a stronger sense of "being there", while the desktop system only provided realism information limited to the screen display, significantly inferior to the VR's immersiveness. Therefore, the VR group was markedly higher than the desktop group in SF and RF.

On the other hand, the data showed no significant differences in CF, DF and EF. The CF evaluates the sense of control and relies on the interaction mode and device capabilities. Both groups used training programs developed from the same engine and tailored to each hardware, thus having satisfactory control experience and comparable CF scores. The DF depends on the experiment isolation extent, which was the same quiet lab for both groups.

The EF requires specific explanation. Theoretically, the desktop experiment imposed smaller physical and mental burden on subjects, while prolonged VR headset use could stress users due to weight. However, no significant EF difference was observed, likely due to the short experiment duration of 5-10 minutes. With longer experiments, accumulated fatigue in the neck caused by prolonged VR headset wear could lead to observable EF differences between the two groups, and relevant studies have given clear results\cite{bienerQuantifyingEffectsWorking2022}.

\subsubsection{Correlation Analysis}

\begin{figure}
	\centering
	\includegraphics[width=\columnwidth]{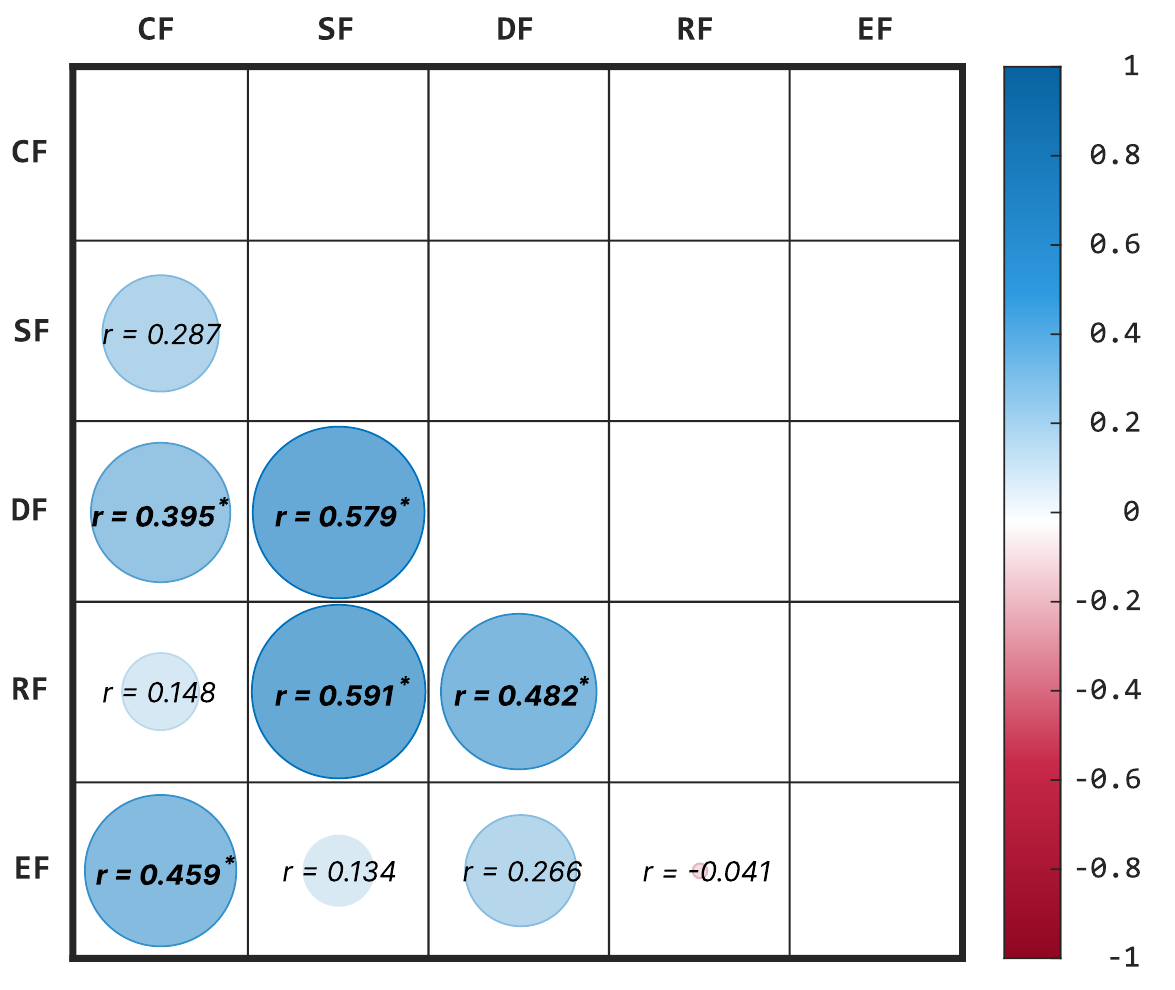}
	\caption{Heatmap of correlations between 5 factors.}
	\label{fig:heatmap}
\end{figure}

Figure \ref{fig:heatmap} depicts the correlation analysis between scale factors, showing significance and correlated elements via colored circles representing Pearson r magnitudes.

This suggests complex, positive inter-factor interactions collectively shaping overall presence. Control factors relate to system operation/control ability. CF correlates more strongly with EF (medium, r$_{(cf,ef)}$=0.459, \textit{p}$<$0.05) than DF (weak, r$_{(cf,df)}$=0.395, \textit{p}$<$0.05), possibly because efficient interaction enhances control and focus by blocking external interference versus the weaker synergy with reducing environmental distraction. Better control may increase presence, relating to human-computer efficiency factors since good interface design aids control and presence.

SF correlates equally with DF and RF (medium, r$_{(sf,df)}$=0.579, r$_{(sf,rf)}$=0.591, \textit{p}$<$0.05). Rich sensory experiences and consistent, filtered information can increase immersion, enhancing SF. Distracting elements like inconsistent realism may reduce SF. High quality stimuli boost realism and consistency, enhancing reality. Sufficient, task-consistent realism may increase presence. Insufficient, inconsistent realism may distract, reducing SF.

DF and RF also have a medium correlation (r$_{(df,rf)}$=0.482, \textit{p}$<$0.05). Distraction factors involve environmental isolation and selective attention. Good isolation and reduced ambient noise augment focus, improving DF. Greater realism may also increase focus, reducing DF.

Collectively, complex control, perception, distraction, etc. interactions shape overall presence.

\subsection{Relationship between presence and learning efficacy}

\begin{figure*}
	\centering
	\includegraphics[width=\textwidth]{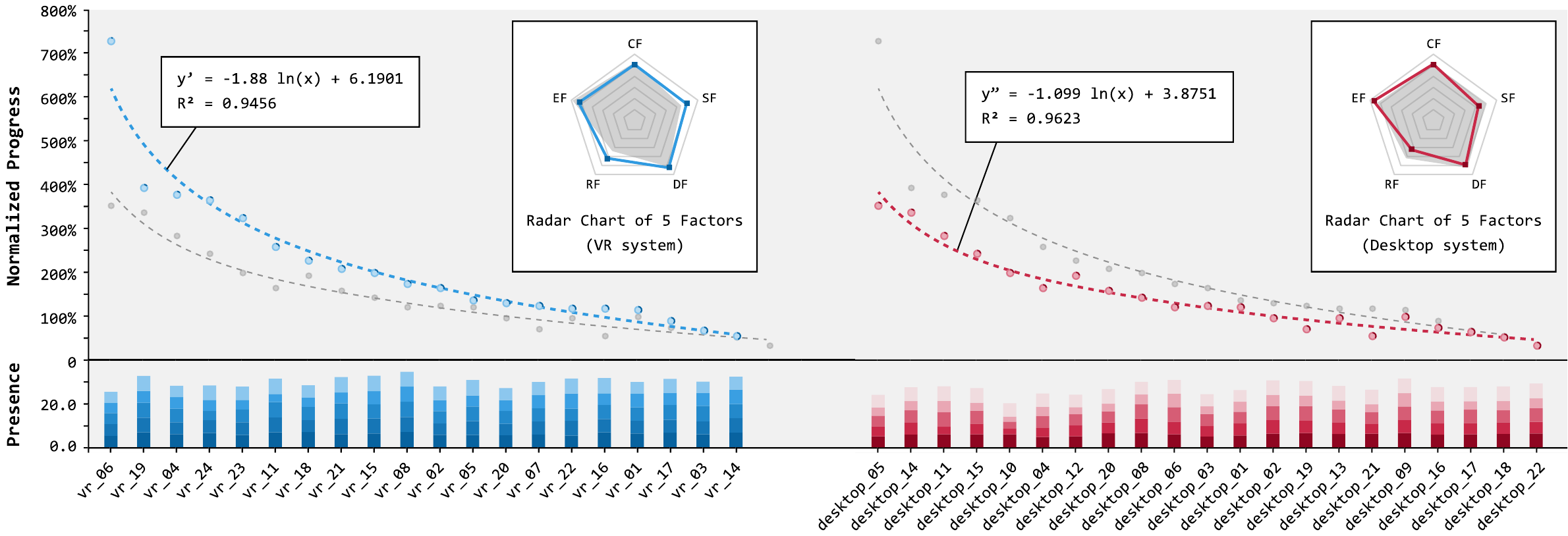}
	\caption{Normalized improvements and scores of 5 factors for each group.}
	\label{fig:np}
\end{figure*}

The presence-efficacy relationship was subsequently explored. Figure \ref{fig:np} plots normalized improvement ratios and presence scores for VR/Desktop groups.

\begin{equation}
	\label{equ:normalized}
	\hat{R}{_{ improvement}} =\frac{N{_{post}} - M{_{pre}}}{M{_{pre}}} \times 100\%
\end{equation}

Normalized improvement ratios (Equation \ref{equ:normalized}) reflect relative improvement over baseline after normalization, unaffected by absolute values and better indicating individuals’ relative gains. Samples are sorted left-to-right by increasing pre-test score. Darker scatter point shading shows the pre-test percentage. VR has higher normalized improvements across all pre-test ranges versus Desktop, more pronounced at lower pre-test scores.

Stacked bar charts represent sample scores on each presence dimension, from bottom to top as CF, SF, DF, RF, EF. Total height indicates total presence. VR has slightly higher overall presence versus Desktop. The radar chart uses trimmed mean group data, showing VR advantages in SF and RF explaining its superior training effect. However, VR has lower human-computer efficiency, likely due to current headsets imposing neck fatigue and motion sickness. Resolving this could further improve VR training.

\section{Conclusion}
Structural cognition training in engineering graphics is vital but challenging. Traditional classroom-based learning with text materials falls short in engaging students and achieving satisfactory outcomes. Physical model training is intuitive but hindered by limited availability and high maintenance costs. This study developed VR and desktop training systems to address these issues, aligning with constructivist learning theory for hands-on practice-based knowledge acquisition.

This study aims to compare and quantify the learning efficacy of immersive VR assembly training and provide insights for its application in engineering education. We analyze three training methods: physical model training, desktop simulation training, and immersive VR simulation training. We also assess presence using Witmer's five factors, comparing presence in the two virtual systems and examining its correlation with learning efficacy.

This study employed rigorous experimental design for reliable results. ANCOVA controlled for individual differences, enabling precise testing of training methods’ effects. Results indicated significantly improved test performance in immersive VR training compared to physical model and desktop virtual training, in line with prior research highlighting VR's superior learning potential. Immersive VR environments enhance task focus, improving learning efficiency. Significant presence differences were observed between VR and desktop systems, particularly in SF, RF, and EF. Among the factors, CF correlates with DF and EF, SF with DF and RF, and DF with RF. Importantly, higher virtual environment presence strongly links to better learning outcomes.

To better understand and explain these results, here are what we believe could be the possible reasons. Firstly, VR allows for safe and repeated simulations of real-world training scenarios, significantly improving learning. Secondly, virtual reality systems enhance learners' perception of task context and concentration, optimizing knowledge acquisition and skill training. Additionally, virtual systems can increase the sense of presence to varying degrees, with stronger immersion leading to higher engagement in learning. This factor underscores the superiority of virtual reality training over traditional methods.

Regarding presence factor correlations, this study also provides insights. The CF-DF and CF-EF correlation is attributed to sensitive human-computer interaction enhancing task focus, even blocking external interference, thus promoting control sense. For instance, research shows natural gesture interaction filters extraneous information to maintain focus\cite{wuExploringFramebasedGesture2022}, increasing control and presence\cite{lingRelationshipIndividualCharacteristics2013}. The SF-DF and SF-RF correlation likely arises since high-quality perceptual stimuli boost environmental realism and consistency, enhancing reality sense. If the virtual assembly system provides realism corresponding to the actual task, presence is easier to experience. Research indicates tangible interaction models can increase virtual world perceptual realism\cite{mcgloinTripleWhammyViolent2015}.

Theoretically, this study supports the use of virtual reality in engineering education, enhancing our understanding of its advantages from a presence perspective. It also explores the relationship between presence and learning efficacy, offering a new perspective on virtual reality training. The practical implications include informing teaching reforms in engineering courses and promoting virtual reality training systems. However, limitations include a small sample size that may not represent all student groups, potential differences between the experimental environment and real teaching settings, and a focus on a simple assembly task without considering more complex scenarios. Future work will expand the research to other training domains for broader applicability.



In summary, this study confirms immersive VR simulation training's advantages in improving student learning, providing evidence supporting virtual reality integration in teaching and guiding subsequent research in this field. As VR technology evolves, immersive VR simulation training will likely play an increasingly vital role in education.

%

%
%
%
%


\bibliographystyle{IEEEtran}
\bibliography{Lib,Software}

\begin{thebibliography}{10}
\providecommand{\url}[1]{#1}
\csname url@samestyle\endcsname
\providecommand{\newblock}{\relax}
\providecommand{\bibinfo}[2]{#2}
\providecommand{\BIBentrySTDinterwordspacing}{\spaceskip=0pt\relax}
\providecommand{\BIBentryALTinterwordstretchfactor}{4}
\providecommand{\BIBentryALTinterwordspacing}{\spaceskip=\fontdimen2\font plus
\BIBentryALTinterwordstretchfactor\fontdimen3\font minus
  \fontdimen4\font\relax}
\providecommand{\BIBforeignlanguage}[2]{{%
\expandafter\ifx\csname l@#1\endcsname\relax
\typeout{** WARNING: IEEEtran.bst: No hyphenation pattern has been}%
\typeout{** loaded for the language `#1'. Using the pattern for}%
\typeout{** the default language instead.}%
\else
\language=\csname l@#1\endcsname
\fi
#2}}
\providecommand{\BIBdecl}{\relax}
\BIBdecl

\bibitem{huoDesignSimulationVehicle2022}
Z.~Huo, X.~Luo, Q.~Wang, V.~Jagota, M.~Jawarneh, and M.~Sharma, ``Design and
  simulation of vehicle vibration test based on virtual reality technology,''
  \emph{Nonlinear Engineering}, vol.~11, no.~1, pp. 500--506, Jan. 2022.

\bibitem{liSynthesizingPersonalizedConstruction2022}
W.~Li, H.~Huang, T.~Solomon, B.~Esmaeili, and L.-F. Yu, ``Synthesizing
  personalized construction safety training scenarios for vr training,''
  \emph{IEEE Transactions on Visualization and Computer Graphics}, vol.~28,
  no.~5, pp. 1993--2002, May 2022.

\bibitem{kickmeier-rustVirtualRealityProfessional2020}
M.~D. {Kickmeier-Rust}, M.~Leitner, and P.~Hann, ``Virtual reality in
  professional training: An example from the field of bank counselling,'' 2020,
  pp. 210--214.

\bibitem{jeonMoreBetterImproving2021}
S.~Jeon, S.~Paik, U.~Yang, P.~C. Shih, and K.~Han, ``The more, the better?
  improving vr firefighting training system with realistic firefighter tools as
  controllers,'' \emph{SENSORS}, vol.~21, no.~21, p. 7193, Nov. 2021.

\bibitem{kennedyImprovingSafetyOutcomes2023}
G.~A.~L. Kennedy, S.~Pedram, and S.~Sanzone, ``Improving safety outcomes
  through medical error reduction via virtual reality-based clinical skills
  training,'' \emph{SAFETY SCIENCE}, vol. 165, p. 106200, Sep. 2023.

\bibitem{luInnovativeVirtualReality2023}
J.~Lu, A.~Leng, Y.~Zhou, W.~Zhou, J.~Luo, X.~Chen, and X.~Qi, ``An innovative
  virtual reality training tool for the pre-hospital treatment of
  cranialmaxillofacial trauma,'' \emph{COMPUTER ASSISTED SURGERY}, vol.~28,
  no.~1, p. 2189047, Dec. 2023.

\bibitem{pedramValidationVRHMDsMedical2023}
S.~Pedram, G.~Kennedy, and S.~Sanzone, ``Toward the validation of vr-hmds for
  medical education: A systematic literature review,'' \emph{VIRTUAL REALITY},
  May 2023.

\bibitem{serranovergelComparativeEvaluationVirtual2020}
R.~Serrano~Vergel, P.~Morillo~Tena, S.~Casas~Yrurzum, and C.~{Cruz-Neira}, ``A
  comparative evaluation of a virtual reality table and a hololens-based
  augmented reality system for anatomy training,'' \emph{IEEE Transactions on
  Human-Machine Systems}, vol.~50, no.~4, pp. 337--348, Aug. 2020.

\bibitem{kullmanVRMRSupporting2019}
K.~Kullman, M.~Ryan, and L.~Trossbach, ``Vr/mr supporting the future of
  defensive cyber operations,'' vol.~52, 2019, pp. 181--186.

\bibitem{ImmerseSaysVR2021}
``Immerse says vr training could help deal with the `great resignation','' Oct.
  2021.

\bibitem{checaImmersiveVirtualrealityComputerassembly2021}
D.~Checa, I.~{Miguel-Alonso}, and A.~Bustillo, ``Immersive virtual-reality
  computer-assembly serious game to enhance autonomous learning,''
  \emph{Virtual Reality}, 2021.

\bibitem{witmerMeasuringPresenceVirtual1998}
B.~G. Witmer and M.~J. Singer, ``Measuring presence in virtual environments: A
  presence questionnaire,'' \emph{Presence: Teleoperators and Virtual
  Environments}, vol.~7, no.~3, pp. 225--240, Jun. 1998.

\bibitem{buttUsingGameBasedVirtual2018}
A.~L. Butt, S.~{Kardong-Edgren}, and A.~Ellertson, ``Using game-based virtual
  reality with haptics for skill acquisition,'' \emph{Clinical Simulation In
  Nursing}, vol.~16, pp. 25--32, Mar. 2018.

\bibitem{sultanExperimentalStudyUsefulness2019}
L.~Sultan, W.~Abuznadah, H.~{Al-Jifree}, M.~A. Khan, B.~Alsaywid, and
  F.~Ashour, ``An experimental study on usefulness of virtual reality
  360\textdegree{} in undergraduate medical education,'' \emph{Advances in
  Medical Education and Practice}, vol.~10, pp. 907--916, Oct. 2019.

\bibitem{zhaoEffectivenessVirtualRealitybased2020}
J.~Zhao, X.~Xu, H.~Jiang, and Y.~Ding, ``The effectiveness of virtual
  reality-based technology on anatomy teaching: A meta-analysis of randomized
  controlled studies,'' \emph{BMC Medical Education}, vol.~20, no.~1, p. 127,
  Apr. 2020.

\bibitem{guedesVirtualRealitySimulator2019}
H.~G. Guedes, Z.~M. C{\^a}mara Costa~Ferreira, L.~{Ribeiro de Sousa Le{\~a}o},
  E.~F. Souza~Montero, J.~P. Otoch, and E.~L. d.~A. Artifon, ``Virtual reality
  simulator versus box-trainer to teach minimally invasive procedures: A
  meta-analysis,'' \emph{International Journal of Surgery}, vol.~61, pp.
  60--68, Jan. 2019.

\bibitem{kandiApplicationVirtualReality2020}
V.~Kandi, P.~Brittle, F.~Castronovo, and C.~Gaedicke, \emph{Application of a
  Virtual Reality Educational Game to Improve Design Review Skills}, Nov. 2020.

\bibitem{sampaioApplicationVirtualReality2014}
A.~Z. Sampaio and O.~P. Martins, ``The application of virtual reality
  technology in the construction of bridge: The cantilever and incremental
  launching methods,'' \emph{Automation in Construction}, vol.~37, pp. 58--67,
  Jan. 2014.

\bibitem{vergaraNewApproachTeaching2017}
D.~Vergara, M.~P. Rubio, and M.~Lorenzo, ``New approach for the teaching of
  concrete compression tests in large groups of engineering students,''
  \emph{Journal of Professional Issues in Engineering Education and Practice},
  vol. 143, no.~2, p. 05016009, Apr. 2017.

\bibitem{chouConstructionVirtualReality1997}
C.~Chou, H.-L. Hsu, and Y.-S. Yao, ``Construction of a virtual reality learning
  environment for teaching structural analysis,'' \emph{Computer Applications
  in Engineering Education}, vol.~5, no.~4, pp. 223--230, 1997.

\bibitem{gomezDevelopmentVirtualEarthquake2018}
D.~Gomez, F.~Guerrero, and P.~Thomson, ``Development of a virtual earthquake
  engineering lab and its impact on education,'' \emph{Dyna (Medellin,
  Colombia)}, vol.~85, Mar. 2018.

\bibitem{mastliInteractiveHighwayConstruction2017}
M.~Mastli and J.~Zhang, ``Interactive highway construction simulation using
  game engine and virtual reality for education and training purpose,'' pp.
  399--406, Jun. 2017.

\bibitem{eswaranChallengesOpportunitiesAR2022}
M.~Eswaran and M.~V. A.~R. Bahubalendruni, ``Challenges and opportunities on
  ar/vr technologies for manufacturing systems in the context of industry 4.0:
  A state of the art review,'' \emph{Journal of Manufacturing Systems},
  vol.~65, pp. 260--278, Oct. 2022.

\bibitem{adasVirtualAugmentedReality2013}
H.~A. Adas, S.~Shetty, and S.~K. Hargrove, ``Virtual and augmented reality
  based assembly design system for personalized learning,'' 2013, pp. 696--702.

\bibitem{winkesMethodEnhancedAssembly2015}
P.~A. Winkes and J.~C. Aurich, ``Method for an enhanced assembly planning
  process with systematic virtual reality inclusion,'' \emph{Procedia CIRP},
  vol.~37, pp. 152--157, 2015.

\bibitem{bharathiInvestigatingImpactInteractive2016}
A.~K. B.~G. Bharathi and C.~S. Tucker, ``Investigating the impact of
  interactive immersive virtual reality environments in enhancing task
  performance in online engineering design activities,'' Jan. 2016.

\bibitem{al-ahmariDevelopmentVirtualManufacturing2016}
A.~M. {Al-Ahmari}, M.~H. Abidi, A.~Ahmad, and S.~Darmoul, ``Development of a
  virtual manufacturing assembly simulation system,'' \emph{Advances in
  Mechanical Engineering}, vol.~8, no.~3, p. 168781401663982, Mar. 2016.

\bibitem{bennettMixedRealityPedagogicalInnovation2021}
V.~G. Bennett, C.~Harteveld, Y.~V. Zastavker, T.~Abdoun, M.~Hossein,
  M.~Omidvar, K.~Wen, and X.~Wirth, ``A mixed-reality pedagogical innovation in
  the reality of a new normal,'' pp. 170--178, May 2021.

\bibitem{abidiAssessmentVirtualRealitybased2019}
M.~H. Abidi, A.~{Al-Ahmari}, A.~Ahmad, W.~Ameen, and H.~Alkhalefah,
  ``Assessment of virtual reality-based manufacturing assembly training
  system,'' \emph{The International Journal of Advanced Manufacturing
  Technology}, vol. 105, no.~9, pp. 3743--3759, Dec. 2019.

\bibitem{hafsiaVirtualRealitySimulator2018}
M.~Hafsia, E.~Monacelli, and H.~Martin, ``Virtual reality simulator for
  construction workers,'' ser. VRIC '18, Apr. 2018, pp. 1--7.

\bibitem{cassolaImmersiveAuthoringVirtual2021}
F.~Cassola, M.~Pinto, D.~Mendes, L.~Morgado, A.~Coelho, and H.~Paredes,
  ``Immersive authoring of virtual reality training,'' 2021, pp. 633--634.

\bibitem{tergasPilotStudySurgical2013}
A.~I. Tergas, S.~B. Sheth, I.~C. Green, R.~Giuntoli, A.~D. Winder, and A.~N.
  Fader, ``Pilot study of surgical training using a virtual robotic surgery
  simulator,'' \emph{Journal of the Society of Laparoendoscopic Surgeons},
  vol.~17, no.~2, pp. 219--226, Apr. 2013.

\bibitem{setarehApplicationVirtualEnvironment2005}
M.~Setareh, D.~A. Bowman, A.~Kalita, M.~Gracey, and J.~Lucas, ``Application of
  a virtual environment system in building sciences education,'' \emph{Journal
  of Architectural Engineering}, vol.~11, no.~4, pp. 165--172, Dec. 2005.

\bibitem{wangTaskComplexityLearning2020}
R.~Wang, R.~Lowe, S.~Newton, and T.~Kocaturk, ``Task complexity and learning
  styles in situated virtual learning environments for construction higher
  education,'' \emph{Automation in Construction}, vol. 113, p. 103148, May
  2020.

\bibitem{becerik-gerberBIMEnabledVirtualCollaborative2012}
A.~M.~A. {Becerik-Gerber}, K.~Ku, and F.~Jazizadeh, ``Bim-enabled virtual and
  collaborative construction engineering and management,'' \emph{Journal of
  Professional Issues in Engineering Education and Practice}, vol. 138, no.~3,
  pp. 234--245, Jul. 2012.

\bibitem{mcmahanEvaluatingDisplayFidelity2012}
R.~P. McMahan, D.~A. Bowman, D.~J. Zielinski, and R.~B. Brady, ``Evaluating
  display fidelity and interaction fidelity in a virtual reality game,''
  \emph{IEEE Transactions on Visualization and Computer Graphics}, vol.~18,
  no.~4, pp. 626--633, Apr. 2012.

\bibitem{duboviNowKnowHow2017}
I.~Dubovi, S.~T. Levy, and E.~Dagan, ``Now i know how! the learning process of
  medication administration among nursing students with non-immersive desktop
  virtual reality simulation,'' \emph{Computers \& Education}, vol. 113, pp.
  16--27, Oct. 2017.

\bibitem{mogliaSystematicReviewVirtual2016}
A.~Moglia, V.~Ferrari, L.~Morelli, M.~Ferrari, F.~Mosca, and A.~Cuschieri, ``A
  systematic review of virtual reality simulators for robot-assisted surgery,''
  \emph{European Urology}, vol.~69, no.~6, pp. 1065--1080, Jun. 2016.

\bibitem{vaughanReviewVirtualReality2016}
N.~Vaughan, V.~N. Dubey, T.~W. Wainwright, and R.~G. Middleton, ``A review of
  virtual reality based training simulators for orthopaedic surgery,''
  \emph{Medical Engineering \& Physics}, vol.~38, no.~2, pp. 59--71, Feb. 2016.

\bibitem{tuzunEffects3DMultiuser2016}
H.~T{\"u}z{\"u}n and F.~{\"O}zdin{\c c}, ``The effects of 3d multi-user virtual
  environments on freshmen university students' conceptual and spatial learning
  and presence in departmental orientation,'' \emph{Computers \& Education},
  vol.~94, pp. 228--240, Mar. 2016.

\bibitem{riegerHowMaximiseSpatial2023b}
M.~B. Rieger and B.~Risch, ``How to maximise spatial presence: Design
  guidelines for a virtual learning environment for school use,'' \emph{IEEE
  Transactions on Visualization and Computer Graphics}, vol.~29, no.~5, pp.
  2517--2526, May 2023.

\bibitem{xieReviewVirtualReality2021}
B.~Xie, H.~Liu, R.~Alghofaili, Y.~Zhang, Y.~Jiang, F.~D. Lobo, C.~Li, W.~Li,
  H.~Huang, M.~Akdere, C.~Mousas, and L.-F. Yu, ``A review on virtual reality
  skill training applications,'' \emph{Frontiers in Virtual Reality}, vol.~2,
  2021.

\bibitem{koumaditisEffectivenessVirtualPhysical2020}
K.~Koumaditis, F.~Chinello, P.~Mitkidis, and S.~Karg, ``Effectiveness of
  virtual versus physical training: The case of assembly tasks, trainer's
  verbal assistance, and task complexity,'' \emph{IEEE Computer Graphics and
  Applications}, vol.~40, no.~5, pp. 41--56, Sep. 2020.

\bibitem{wintherDesignEvaluationVR2020}
F.~Winther, L.~Ravindran, K.~Svendsen, and T.~Feuchtner, ``Design and
  evaluation of a vr training simulation for pump maintenance based on a use
  case at grundfos,'' 2020, pp. 738--746.

\bibitem{chenApplicationAugmentedReality2011}
H.~Chen, K.~Feng, C.~Mo, S.~Cheng, Z.~Guo, and Y.~Huang, ``Application of
  augmented reality in engineering graphics education,'' vol.~2, Dec. 2011, pp.
  362--365.

\bibitem{kolbExperientialLearningExperience2015}
D.~A. Kolb, \emph{Experiential Learning: Experience as the Source of Learning
  and Development}, second edition~ed., 2015.

\bibitem{brownSituatedCognitionCulture1989}
J.~S. Brown and A.~Others, ``Situated cognition and the culture of learning,''
  \emph{Educational Researcher}, vol.~18, no.~1, pp. 32--42, 1989.

\bibitem{winVisualEffectsCreation2021}
L.~L. Win, A.~A. Faieza, A.~A. Hairuddin, L.~N. Abdullah, H.~J. Yap, H.~Saito,
  and S.~Hisham, ``Visual effects creation to improve user performance in fully
  immersive virtual environment of automobile engine assembly,'' \emph{IOP
  Conference Series: Materials Science and Engineering}, vol. 1109, no.~1, p.
  012038, Mar. 2021.

\bibitem{rostamiMetaverseImplementingAdvanced2022}
S.~Rostami and M.~Maier, ``The metaverse and beyond: Implementing advanced
  multiverse realms with smart wearables,'' \emph{Ieee Access}, vol.~10, pp.
  110\,796--110\,806, 2022.

\bibitem{yuanSimplifiedTessellatedMesh2016}
Y.~Yuan, R.~Wang, J.~Huang, Y.~Jia, and H.~Bao, ``Simplified and tessellated
  mesh for realtime high quality rendering,'' \emph{Computers \& Graphics},
  vol.~54, pp. 135--144, Feb. 2016.

\bibitem{yangDeepFaceVideo2023}
W.~Yang, Z.~Chen, C.~Chen, G.~Chen, and K.-Y.~K. Wong, ``Deep face video
  inpainting via uv mapping,'' \emph{Ieee Transactions on Image Processing},
  vol.~32, pp. 1145--1157, 2023.

\bibitem{hendrickxRealisticRenderingVirtual2011}
Q.~Hendrickx, K.~Van~Golen, Z.~K. Shao, and M.~Driel, ``Realistic rendering of
  virtual worlds,'' 2011.

\bibitem{mayerComputerGamesLearning2014}
R.~E. Mayer, \emph{Computer Games for Learning: An Evidence-Based Approach},
  2014.

\bibitem{jamovi_2022}
{The jamovi project}, ``{jamovi},'' \url{https://www.jamovi.org}, 2022,
  computer software.

\bibitem{R_2021}
{R Core Team}, ``{R}: A language and environment for statistical computing,''
  \url{https://cran.r-project.org}, 2021, computer software.

\bibitem{car_package}
J.~Fox and S.~Weisberg, ``{car}: Companion to applied regression,''
  \url{https://cran.r-project.org/package=car}, 2020, r package.

\bibitem{MATLAB_2022b}
{MathWorks}, ``{MATLAB},''
  \url{https://www.mathworks.com/products/matlab.html}, R2022b, computer
  software.

\bibitem{lamImpactCompetitivenessSalespeople2012}
L.~W. Lam, ``Impact of competitiveness on salespeople's commitment and
  performance,'' \emph{Journal of Business Research}, vol.~65, no.~9, pp.
  1328--1334, 2012.

\bibitem{fornellEvaluatingStructuralEquation1981}
C.~Fornell and D.~F. Larcker, ``Evaluating structural equation models with
  unobservable variables and measurement error,'' \emph{Journal of Marketing
  Research}, vol.~18, no.~1, pp. 39--50, 1981.

\bibitem{bienerQuantifyingEffectsWorking2022}
V.~Biener, S.~Kalamkar, N.~Nouri, E.~Ofek, M.~Pahud, J.~J. Dudley, J.~Hu, P.~O.
  Kristensson, M.~Weerasinghe, K.~{\v C}. Pucihar, M.~Kljun, S.~Streuber, and
  J.~Grubert, ``Quantifying the effects of working in vr for one week,''
  \emph{IEEE Transactions on Visualization and Computer Graphics}, vol.~28,
  no.~11, pp. 3810--3820, Nov. 2022.

\bibitem{wuExploringFramebasedGesture2022}
H.~Wu, S.~Fu, L.~Yang, and X.~L. Zhang, ``Exploring frame-based gesture design
  for immersive vr shopping environments,'' \emph{Behaviour \& Information
  Technology}, vol.~41, no.~1, pp. 96--117, Jan. 2022.

\bibitem{lingRelationshipIndividualCharacteristics2013}
Y.~Ling, H.~T. Nefs, W.-P. Brinkman, C.~Qu, and I.~Heynderickx, ``The
  relationship between individual characteristics and experienced presence,''
  \emph{Computers in Human Behavior}, vol.~29, no.~4, pp. 1519--1530, Jul.
  2013.

\bibitem{mcgloinTripleWhammyViolent2015}
R.~McGloin, K.~M. Farrar, and J.~Fishlock, ``Triple whammy! violent games and
  violent controllers: Investigating the use of realistic gun controllers on
  perceptions of realism, immersion, and outcome aggression,'' \emph{Journal of
  Communication}, vol.~65, no.~2, pp. 280--299, 2015.

\end{thebibliography}

%
%


\begin{IEEEbiography}[{\includegraphics[width=1in,height=1.25in,clip,keepaspectratio]{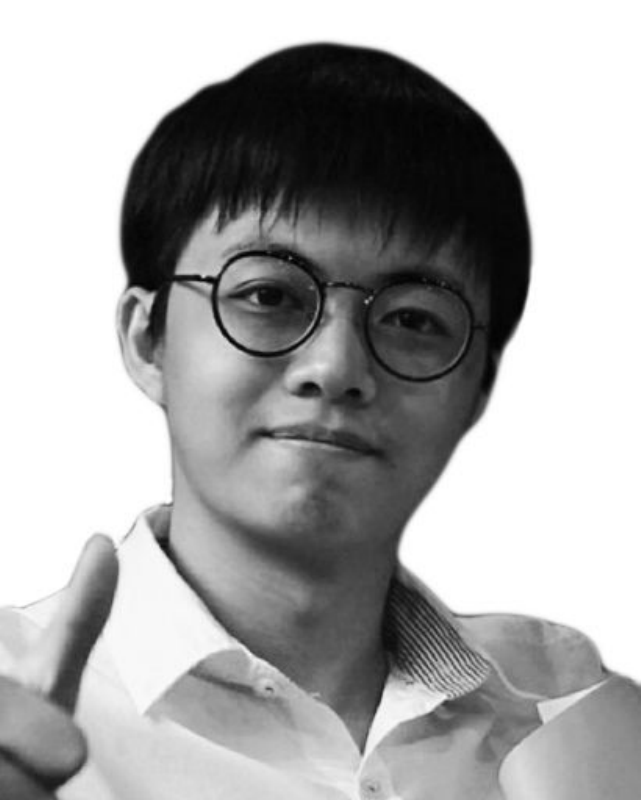}}]{Weichao Lin} was born in Guangdong, China in 1999. received the B.S. degree in Information and Interaction Design from South China University of Technology (SCUT), Guangzhou, China, in 2022. He is currently working toward the M.S. degree in Industrial Design Engineering with the School of Design, SCUT.

	He currently serves as a teaching assistant for the Engineering Drawing course. His research interests include 3D interface, virtual reality, interaction design, and user experience.
\end{IEEEbiography}

\begin{IEEEbiography}[{\includegraphics[width=1in,height=1.25in,clip,keepaspectratio]{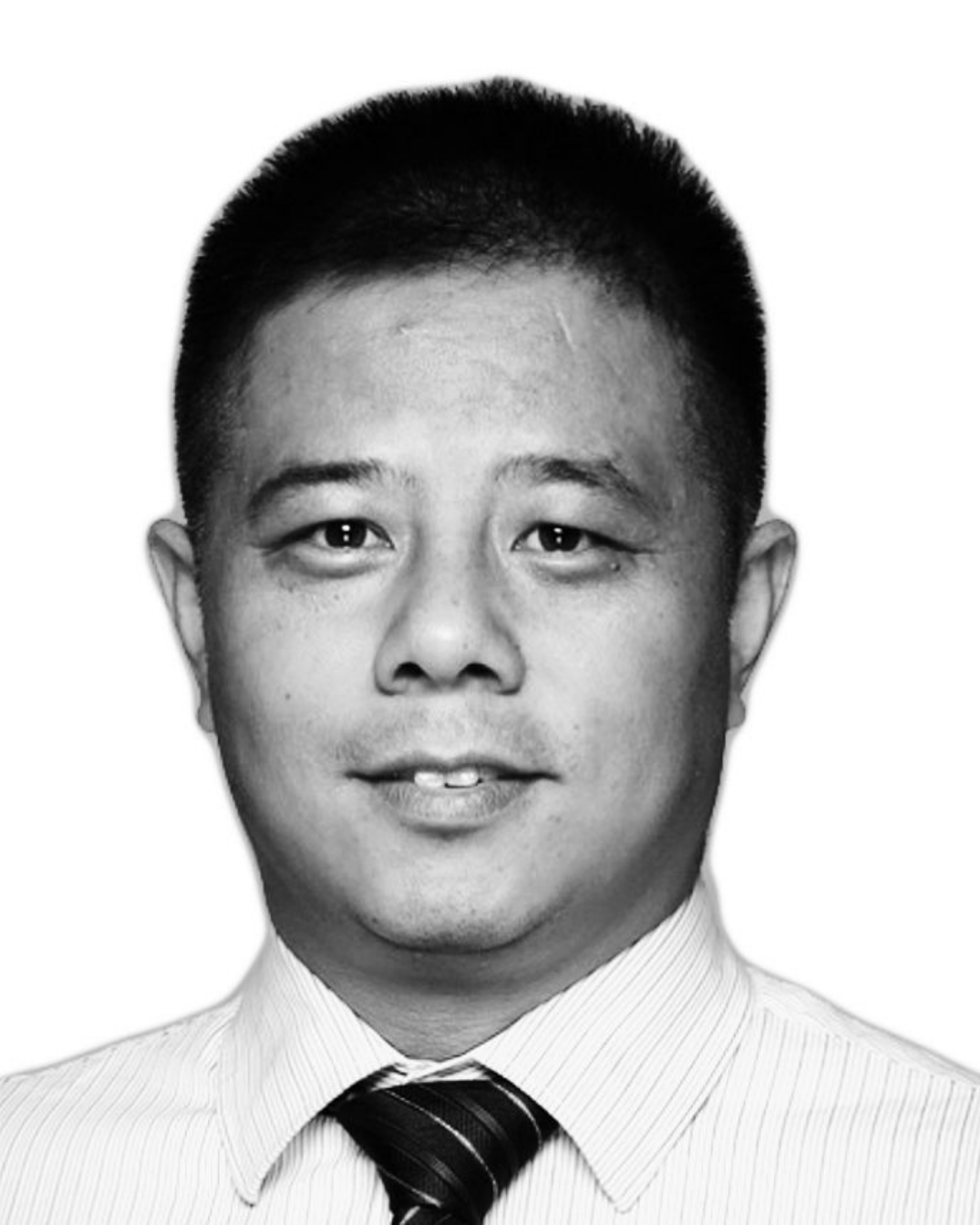}}]{Liang Chen} was born in Liaoning, China in 1975. He received the M.S. degree in Mechanical Design and Theory, and the Ph.D. degrees in Mechanical Manufacture and Automation from South China University of Technology (SCUT), Guangzhou, China, in 2004 and 2014, respectively.

He is currently a Professor with the Information and Interaction Design Department, SCUT China. He is also member of the China Graphics Society. His research interests include spatial ability training, virtual reality, and interaction design.

\end{IEEEbiography}

\begin{IEEEbiography}[{\includegraphics[width=1in,height=1.25in,clip,keepaspectratio]{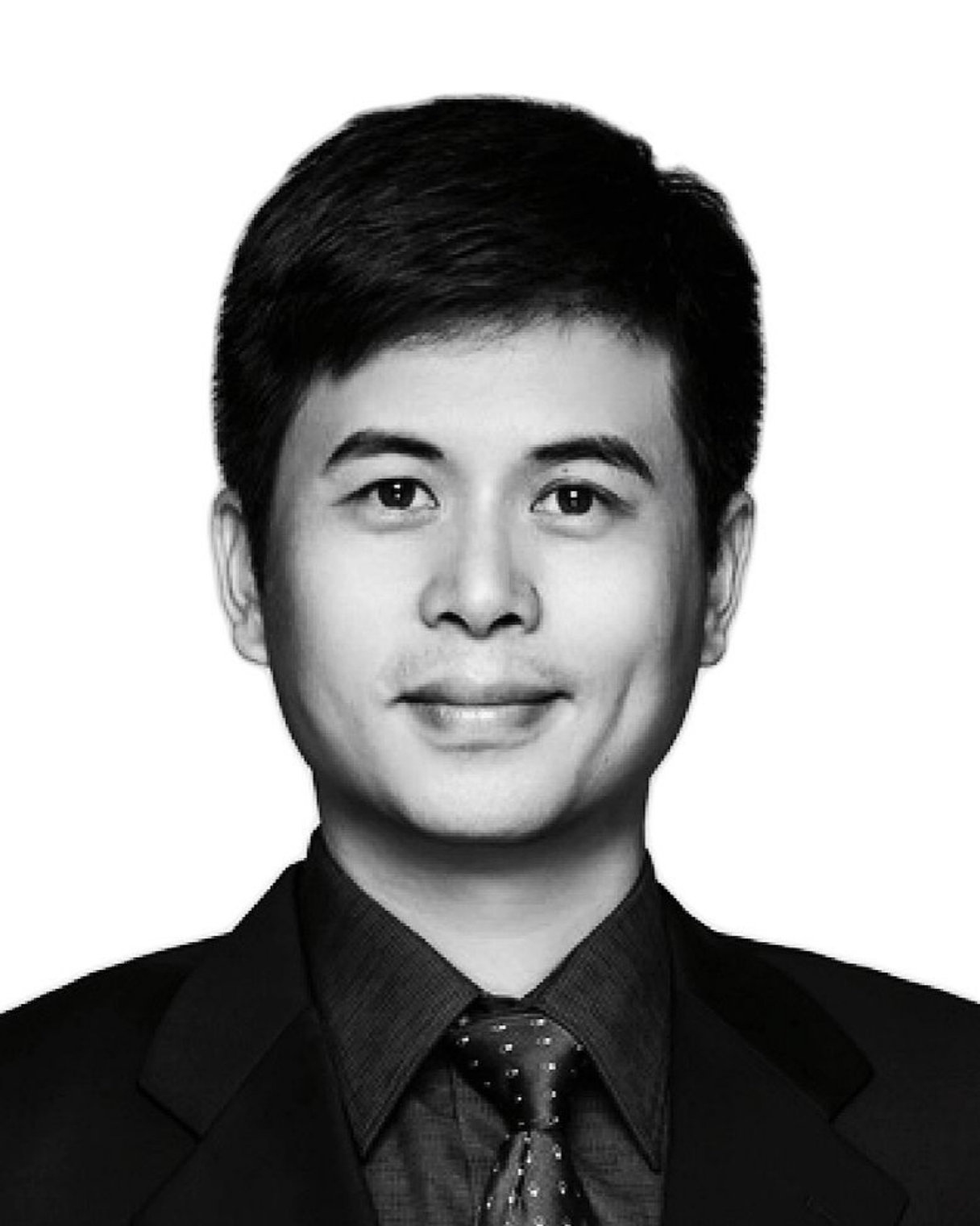}}]{Wei Xiong} was born in Jiangxi, China in1979. He received the B.S. degree in Industrial Equipment and Control Engineering, the M.S. degree in Mechanical Design, and the Ph.D. degree in Mechanical Manufacture and Automation from South China University of Technology (SCUT), Guangzhou, China, in 2002, 2006, and 2017, respectively.

	He is currenty an Professor with the School of Design, SCUT. His expertise is in virtual reality and augmented reality.
\end{IEEEbiography}

\begin{IEEEbiography}[{\includegraphics[width=1in,height=1.25in,clip,keepaspectratio]{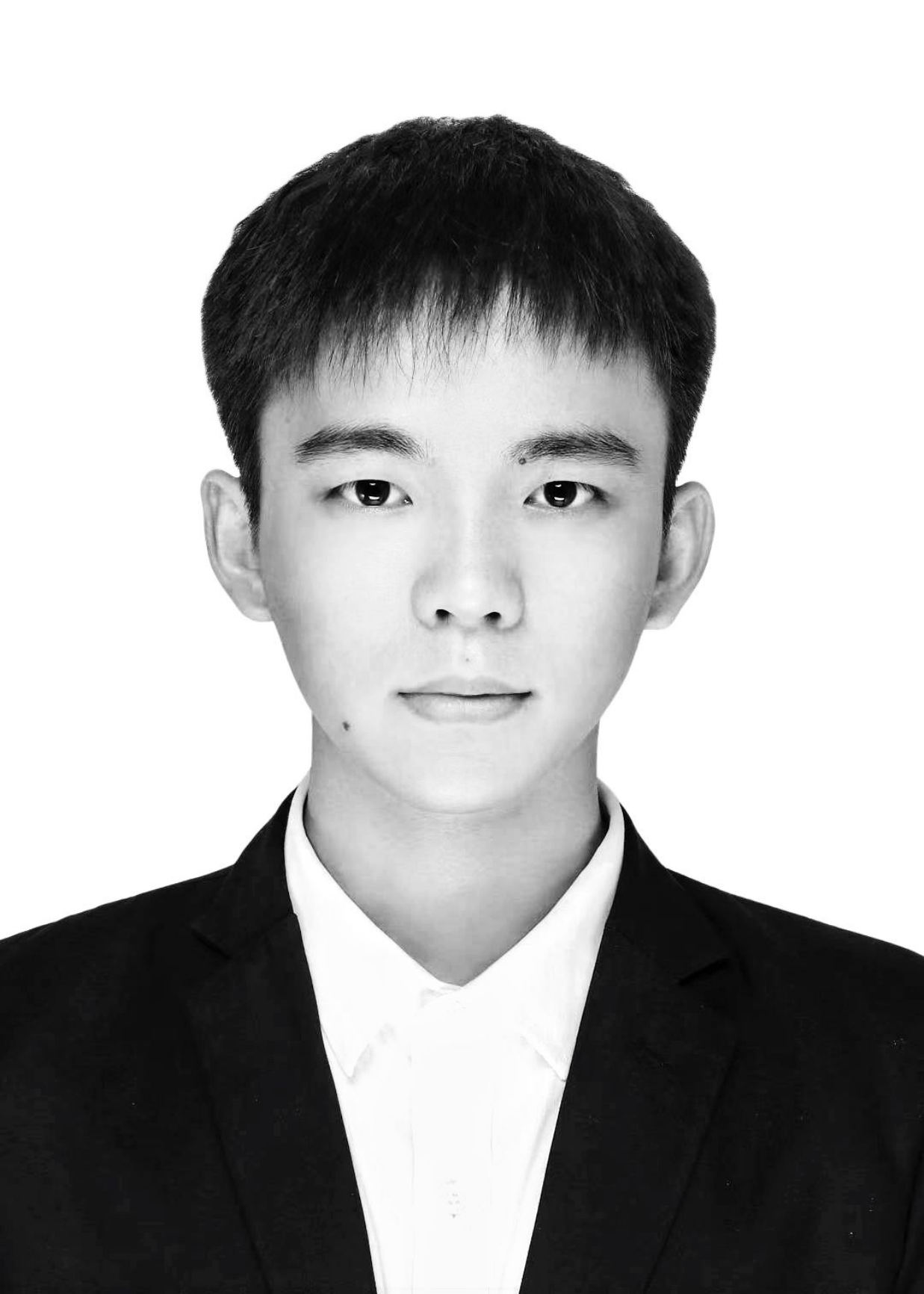}}]{Kang Ran} was born in Chongqing, China, in 1999. He earned his bachelor's degree in Information and Interaction Design from South China University of Technology (SCUT) in 2022. He then continued his studies at the Graduate School of Design at SCUT.

	He currently serves as a teaching assistant for the Engineering Drawing course. His research interests include virtual reality, interaction design, serious games, and artificial intelligence.
\end{IEEEbiography}

\begin{IEEEbiography}[{\includegraphics[width=1in,height=1.25in,clip,keepaspectratio]{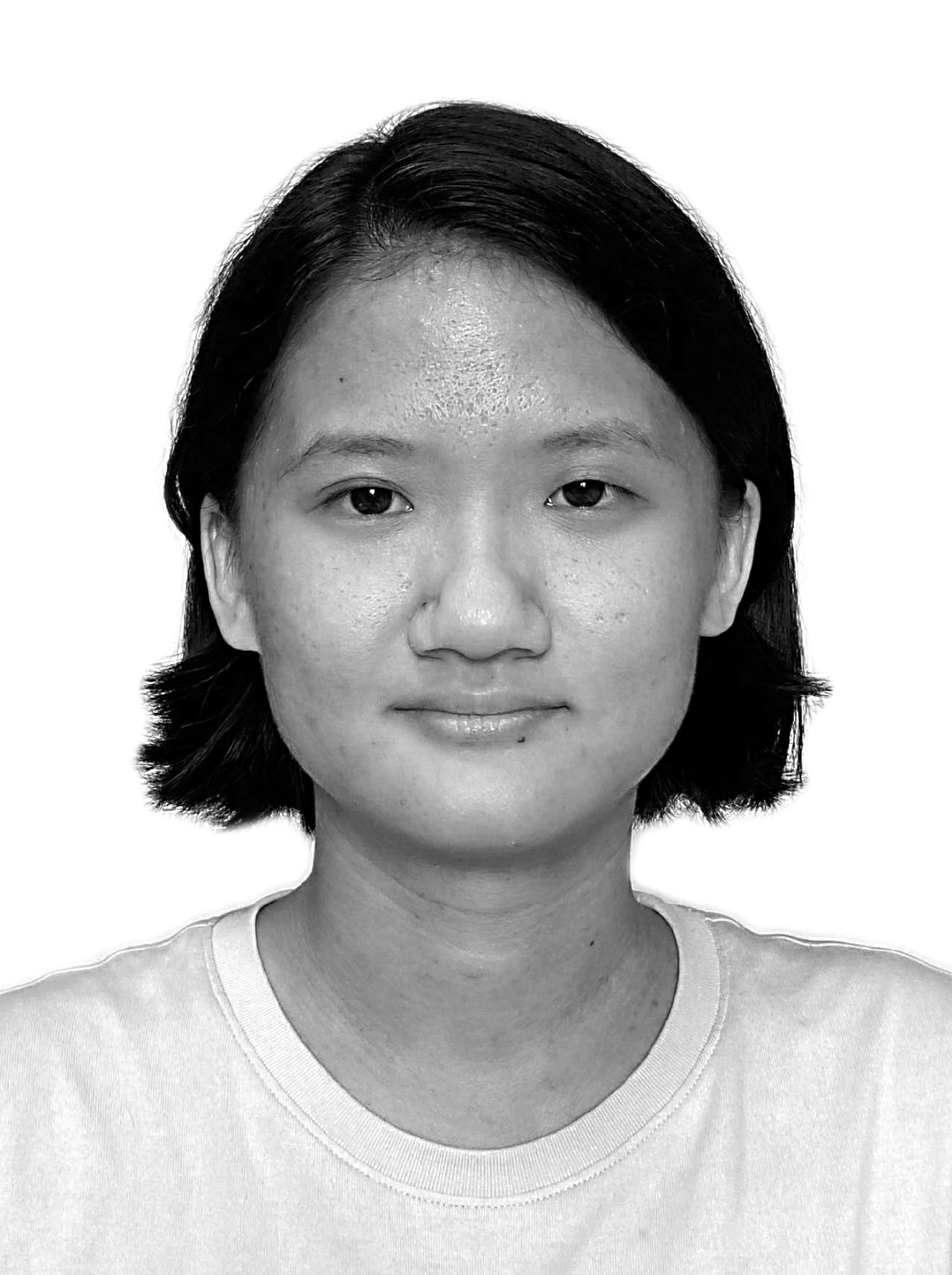}}]{Anlan Fan} Anlan Fan was born in Guangdong, China, in 1997. She has received the Bachelor degree in Computer Science from Beijing Institute of Technology (BIT), Beijing, China, in 2020. She is currently working towards the M.S. degree with the South China University of Technology (SCUT).

	She has been working with Advanced Computing in the School of Computer Science and Technology of the Australian Nation University as an exchange student. Her research focuses on analyzing user statistics of products for intelligent design.
\end{IEEEbiography}

\vspace{11pt}


\vfill


\end{document}